\documentclass[11pt,a4paper]{article}
\usepackage{jheppub}
\usepackage{amsthm}
\usepackage{mathtools}
\usepackage{physics}
\usepackage{array}
\usepackage{enumitem}
\usepackage{bm}
\usepackage{hyperref}
\usepackage[capitalize]{cleveref}
\usepackage{graphicx}
\usepackage{xcolor}
\newcommand{\Q}{\mathbb{Q}}
\newcommand{\C}{\mathbb{C}}

\newcommand{\Z}{\mathbb{Z}}
\newcommand{\K}{\Bbbk}
\newcommand{\PP}{\mathbb{P}}
\title{Anomaly cancellation for two $U(1)$ factors}
\author[a]{Ben Gripaios}
\author[b]{and Khoi {Le Nguyen Nguyen}}

\affiliation[a]{Cavendish Laboratory, University of Cambridge, \\ J.J. Thompson Avenue, Cambridge, CB3 0US, U.K.}

\affiliation[b]{DAMTP, University of Cambridge, \\ Wilberforce Road, Cambridge, CB3 0WA, U.K.}

\emailAdd{gripaios@hep.phy.cam.ac.uk}
\emailAdd{kl518@cam.ac.uk}

\abstract{We show that solving the abelian part of the local anomaly cancellation conditions for a 4-d gauge theory whose gauge Lie algebra has an abelian summand with rank $K \geq 1$ is equivalent to the problem in algebraic/arithmetic geometry of finding $(K-1)$-dimensional projective linear subspaces of a cubic hypersurface over the rational numbers, where the cubic is determined by the data of 	the semisimple summand and the representation thereof carried by the Weyl fermions. We then use this reformulation to solve a variety of examples with rank 2.
	
The simplest non-trivial example from physics, namely gauge Lie group $U(1)^2$ and six fermions, nevertheless has a rich (and well-studied) geometry: it corresponds to the Fano variety of lines in the Segre cubic primal threefold. This is a surface with 15 irreducible components that are planes (which correspond to lines lying in the 15 planes in the Segre cubic primal and which give rise to non-chiral fermions) and 6 components that are split del Pezzo surfaces of degree 5 (which give rise to chiral fermions). These components are all rational varieties, enabling all solutions to the anomaly cancellation conditions to be parametrized and their collective properties (e.g. their topology and asymptotic distribution) to be described.} 

\begin{document}
	
	\maketitle
	
	\section{Introduction}\label{sec:intro}
	This paper is part of a series of works \cite{Allanach_2020_geometricu1,Gripaios_2024_irreducible,Gripaios_2025_products,Gripaios_2026_u1} that 
	systematically address the problem of local anomaly cancellation in gauge quantum field theories in four spacetime dimensions by connecting it to arithmetic geometry. Until recently, this was considered a hopeless endeavour, because it boils down to solving multiple cubic polynomial equations in integers that label the representations of the gauge Lie algebra (either the charges for the abelian summand or the Dynkin labels of representations of the semisimple summand).
	
	To set the scene, the first real breakthrough in this direction came with \cite{Costa_2019}, where it was realized that given two solutions to the anomaly cancellation equations of the simplest theory with gauge Lie algebra $\mathfrak{u}_1$ (and an arbitrary number of fermions), one is able to construct a third. Even better, this often enables one to construct chiral solutions (which are the ones of most interest for low-energy physics) from non-chiral (a.k.a. vector-like) ones (which are easy to find, but of less interest).
	
	It was subsequently explained in \cite{Allanach_2020_geometricu1} that this construction is nothing but the method of secants and tangents, known to geometers since antiquity. In modern language, the fact that the 
	anomaly cancellation equations are homogeneous equations in the ring of integers means that they define a projective variety (in fact, a cubic hypersurface) over the field of rational numbers, and the secant and tangent construction allows one to show that this variety is a rational variety in the $\mathfrak{u}_1$ case when the number of fermions exceeds four. This fact enables all solutions to be found and their collective features to be studied using the powerful tools of algebraic geometry.  
	(To give an explicit example, in the first non-trivial case of five fermions, the variety that results is the famous Clebsch diagonal cubic surface, the non-chiral solutions correspond to points on the 15 rational lines on the surface, and applying the secant and tangent construction to the points on any pair of skew lines defines a birational map from $\mathbb{Q}P^2$ to the variety, exhibiting it as a rational variety.) From there, it follows immediately that the number of chiral solutions is infinite and that they overwhelm the non-chiral solutions, in the precise sense that the former are dense in, e.g., the Zariski topology, while the latter are not \cite{Gripaios_2024_irreducible}.
	
	Remarkably, the very same varieties arise in the problem of finding the anomaly-free irreducible representations of semisimple Lie algebras \cite{Gripaios_2024_irreducible, Gripaios_2025_products}, an apparent coincidence whose significance has yet to be understood, but whose existence again allows for a complete solution of the problem.
	
	The next step was to attack \cite{Gripaios_2026_u1} the case of a gauge Lie algebra with a single $\mathfrak{u}_1$ summand accompanied by a semisimple summand, such as $\mathfrak{su}_2\oplus\mathfrak{u}_1$ or the $\mathfrak{su}_3\oplus \mathfrak{su}_2\oplus\mathfrak{u}_1$ of the Standard Model. Here we get projective varieties that are still cubic hypersurfaces, but which are no longer rational varieties. Nevertheless, they are (typically) unirational varieties (meaning they admit a dominant rational map from some projective space), enabling one to find infinitely many solutions starting from just a single solution (which again can be non-chiral, so easy to find).
	
	In this paper, we study a problem which appears to be much more formidable, namely the case where the gauge Lie algebra has an abelian summand of higher rank, i.e. multiple $\mathfrak{u}_1$ summands, perhaps accompanied by a semisimple summand. Now there are many more polynomial equations to be solved (see \cref{eqn:anomaly_grav_grav_u1,eqn:anomaly_simple_simple_u1,eqn:anomaly_u1_u1_u1}) and they determine a projective variety that is neither cubic nor a hypersurface, in general. Nevertheless, we will see that in certain cases one can make progress and indeed find a complete solution. (Somewhat counter-intuitively, we will see in \cref{sec:cubic_surface} that there are even cases where the problem with a single $\mathfrak{u}_1$ and a fixed semisimple summand cannot be solved with current methods, but the problem with two $\mathfrak{u}_1$ summands can.) While all of our examples feature just two $\mathfrak{u}_1$ summands, the underlying formalism applies to any number of summands. 
	
	To see that cases with multiple $\mathfrak{u}_1$ summands are not completely hopeless, consider the simplest case of the gauge Lie algebra $\mathfrak{u}_1\oplus \mathfrak{u}_1$, i.e. the gauge Lie group $U(1)\times U(1)$. Here, \cite{Costa_2020} showed (by pure algebra) that there are no chiral solutions with up to five fermions, while for six fermions, given any solution to the $\mathfrak{u}_1$ case (for which all solutions can be found, as described above), one can construct six solutions for the charges under the other $\mathfrak{u}_1$ and that in this way one can get a general solution.
	
	Unfortunately, this approach of using solutions for the $\mathfrak{u}_1$ case to find solutions for
	the case with multiple $\mathfrak{u}_1$ summands does not work in general, as we shall show in the next Section; nor does it allow us to gain much geometric insight, even in this simple example. Thus, we are led here to propose a different approach, which again involves
	systematically applying existing tools of algebraic geometry. The key insight (which was already pointed out in the case of two-dimensional gauge theories in \cite{Camp_2024}) is that whereas solutions to the case with a single $\mathfrak{u}_1$ summand correspond to points on some cubic hypersurface, solutions to the case with $K$ $\mathfrak{u}_1$ summands correspond (in a way we make explicit in \cref{sec:alggeo}) to $(K-1)$-dimensional projective linear subspaces on the {\em same} cubic hypersurface. This allows us to connect with another well-studied subject in 
	algebraic geometry, namely that of Fano varieties of linear subspaces of cubic hypersurfaces, which we call Fano varieties for short.\footnote{We warn the reader that there is a distinct class of varieties in algebraic geometry that also go by the name of Fano varieties.}
	
	We begin with the simplest non-trivial class of such Fano varieties, which are the varieties of lines in cubic surfaces. These are zero-dimensional varieties and famously consist of 27 points, at least when we work over an algebraically-closed field, such as the complex numbers. However, this number may be lower when we work, as we must, over the rational numbers; as we will see in \cref{sec:u1u1} this leads to interesting examples in physics already in the case of a gauge Lie algebra $\mathfrak{su}_2\oplus\mathfrak{u}_1\oplus\mathfrak{u}_1$ with six fermions. ({\em A priori} there is an even simpler case, namely $\mathfrak{u}_1\oplus\mathfrak{u}_1$ with five\footnote{Because of the presence of mixed anomalies between abelian and semisimple factors in four dimensions, which lead to linear constraints on the $\mathfrak{u}_1$ charges, the dimension of the projective variety depends on both the number of fermions $I$ and the number of simple factors $J$ in the gauge Lie algebra, as $I-J-3$.} fermions, but there we show that all 15 rational lines lead to solutions which are boring in the sense that they all have non-chiral charges, reproducing the result of \cite{Costa_2020}, but via a geometric approach.)
	
	The next simplest class of such varieties are the Fano varieties of lines in cubic 3-folds, which are themselves two-dimensional surfaces, typically. This includes the aforementioned case of gauge Lie algebra $\mathfrak{u}_1\oplus\mathfrak{u}_1$ with six fermions studied in \cite{Costa_2020}, where our geometric approach allows a great deal more insight. The corresponding cubic 3-fold is again a much-studied object in algebraic geometry (at least over $\mathbb{C}$) where it goes by the name of the Segre cubic primal \cite{Segre_1886,Castelnuovo_1891,Richmond_1902,Manin_1986,Dolgachev_2012,Dolgachev_2016,Ciliberto_2024_gromov,Ciliberto_2024_modular}. As a result, most of our work in this case is reduced to: {\em (i)} ringing the changes that occur when we instead work over the rationals; {\em (ii)} giving, for the benefit of physicist readers, the results
	in a way that requires as little knowledge of algebraic geometry as possible; and {\em (iii)} providing explicit formul\ae\ that can be used in physics applications.
	
	As we describe in \cref{sec:u1u1}, in this case the Fano variety of lines has degree 45 and is reducible, having
	15 irreducible components of degree 1 and 6 components of degree 5. 
	The former are of course isomorphic to projective planes and correspond to lines in each of the 15 planes in the Segre cubic primal; these solutions all correspond to solutions of the anomaly cancellation equations with non-chiral charges. The remaining six components correspond to chiral solutions and each is isomorphic to a del Pezzo surface of degree five, {\em i.e.} a blow-up of the projective plane in four points in general position.\footnote{One subtlety in working over $\Q$ as opposed to the usual case of $\C$ is that, over a non-algebraically closed field, there are now multiple isomorphism classes of del Pezzo surfaces of degree five up to isomorphism \cite{Skorobogatov_2001,Poonen_2017}. However, over a perfect field such as $\Q$, there is exactly one isomorphism class of these surfaces, whose automorphism group is $S_5$; such a surface is a blow-up of the projective plane $\Q P^2$ at four rational points \cite{Boitrel_2025}, otherwise known as the \emph{split} del Pezzo surface.}
	
	With this knowledge, the results of \cite{Costa_2020} can be easily recovered, but one can go much further.
	Indeed, it will turn out that each of the six families of solutions in \cite{Costa_2020} corresponds to one of the six components isomorphic to a del Pezzo surface. So the six families are seen to be equivalent (in the sense of parametrizing isomorphic varieties (and indeed permuting the six charges gives a transitive action on the six components, with stabilizer $S_5$). Moreover, each of the 21 components of the Fano variety of lines is manifestly seen to be a rational surface (this is a tautology for the 15 planes and follows immediately from the blow-up property for the 6 del Pezzo components), meaning that they admit a birational map from $\mathbb{Q}P^2$; any such map enables us to parametrize nearly all (a dense open set, to be precise) of the solutions in the most efficient way possible\footnote{By way of comparison, the parametrization of solutions given in \cite{Costa_2020} requires six integer parameters, while in our approach essentially just two rational parameters are required, corresponding to the fact that the Fano variety is two-dimensional.} and study their geometrical, topological, and arithmetic properties.  We will give examples of such maps in \cref{sec:u1u1}. One particular map is derived from a different birational map, namely one from $\mathbb{Q}P^3$ to the underlying Segre cubic primal 3-fold given over a century ago by Richmond \cite{Richmond_1902}. This map is defined everywhere on $\mathbb{Q}P^3$ apart from five points and has a number of nice properties that make the geometry transparent and help in
	answering questions of relevance to physics. For example, as well as giving an efficient parametrization of solutions,  it makes it easy to solve the problem of, given a set of anomaly-free charges under one $\mathfrak{u}_1$,  finding all possible values for anomaly-free charges under the other $\mathfrak{u}_1$, recovering the results of \cite{Costa_2020} in a particularly transparent way. To wit, given a sixth point in $\mathbb{Q}P^3$, which via Richmond's map defines a point on the Segre cubic primal and thus a solution to the case with one 
	$\mathfrak{u}_1$, one can obtain six lines in the cubic through it as follows. Five lines are obtained as the images under Richmond's map of the lines through this sixth point and the five points in $\mathbb{Q}P^3$ at which the map is undefined, while the sixth line is the image of 
	the unique twisted cubic through all six points in $\mathbb{Q}P^3$. The points on these six lines correspond to the allowed values of the charges under the other $\mathfrak{u}_1$. 
	
	Our geometric formulation also allows us to explain another mysterious phenomenon observed in \cite{Costa_2020}, namely the existence of solutions which are chiral with respect to the full gauge Lie algebra $\mathfrak{u}_1 \oplus \mathfrak{u}_1$, but non-chiral with respect to either $\mathfrak{u}_1$. These arise because the chiral lines can intersect planes in the cubic, whose points are non-chiral. As we show, a generic line intersects five distinct planes, and so there are 10 choices of pairs of points on such a line (namely where each point lies on a distinct plane) that give rise to such solutions.
	
	The outline is as follows. In the next Section, we describe the algebro-geometric fomulation and show why the method of finding lines through points used by \cite{Costa_2020} in the pure $\mathfrak{u}_1 \oplus \mathfrak{u}_1$ case does not work in general. We solve a variety examples from physics with a $\mathfrak{u}_1 \oplus \mathfrak{u}_1$ summand that lead to cubic surfaces in \cref{sec:cubic_surface} and solve the pure $\mathfrak{u}_1 \oplus \mathfrak{u}_1$ example that leads to a cubic 3-fold in \cref{sec:u1u1}. In \cref{sec:otherex} we sketch some qualitative features of other classes of examples with a $\mathfrak{u}_1 \oplus \mathfrak{u}_1$ summand.
	\section{Formulation}\label{sec:alggeo}
	In this Section, we recall the familiar anomaly cancellation conditions and describe the connection to algebraic geometry. We will also discuss various subtleties regarding when solutions to anomaly cancellation conditions should be regarded as physically acceptable, or physically equivalent, and explain why the strategy of finding lines through points does not work in general.
	
	A gauge quantum field theory in four spacetime dimensions has a gauge Lie algebra $\mathfrak{g}$ that is a sum of $J$ simple algebras $\mathfrak{g}_j$, $j\in\{1,\dots,J\}$ (whose sum is the \emph{semisimple summand}) and $K$ copies of $\mathfrak{u}_1$ (whose sum is the \emph{abelian summand}). Its chiral fermions carry a representation of $\mathfrak{g}$ that is $(I-1)$-fold reducible, meaning we have $I$ 
	chiral fermions, each with the charge under the $k$-th $\mathfrak{u}_1$ summand denoted by $x_{ki} \in \mathbb{Z}$, with ${i\in\{0,\dots,I-1\}}$\footnote{The index begins with 0 so as to match the usual conventions for homogeneous co-ordinates in projective geometry.} and $k\in\{1,\dots,K\}$. Supposing that the purely semisimple part of the anomaly cancels as it must,\footnote{The algebro-geometric approach to this is treated in \cite{Gripaios_2024_irreducible,Gripaios_2025_products}.} the remaining local anomaly cancellation conditions are given by polynomial equations in $x_{ki}$:
	\begin{align}
		\sum_{i=0}^{I-1} T_i^{j}\left(\prod_{\substack{j'=1 \\ j'\neq j}}^{J} D_i^{j'}\right) x_{ki} &= 0,&&j\in\{1,\dots,J\},k\in\{1,\dots,K\}, \label{eqn:anomaly_simple_simple_u1}\\
		\sum_{i=0}^{I-1}\left(\prod_{j=1}^{J} D_i^j\right) x_{ki} &= 0,&&k\in\{1,\dots,K\}, \label{eqn:anomaly_grav_grav_u1}\\
		\sum_{i=0}^{I-1} \left(\prod_{j=1}^{J} D_i^j\right) x_{ki}x_{k'i}x_{k''i} &= 0,&&k,k',k''\in\{1,\dots,K\}. \label{eqn:anomaly_u1_u1_u1}
	\end{align}
	Here, $D_i^j \in \mathbb{N}$ and $T_i^j \in \mathbb{N}$ are, respectively, the dimension and Dynkin index\footnote{See \cite{Gripaios_2026_asymp} for a discussion of the integer normalization of the Dynkin index.} of the irreducible representation labelled by $i$ restricted to the simple summand $j$. As is well-known, the $JK$ equations \cref{eqn:anomaly_simple_simple_u1} come from consideration of triangular one-loop Feynman diagrams with one abelian gauge boson leg and two simple gauge boson legs, while the $K$ equations in \cref{eqn:anomaly_grav_grav_u1} come from {replacing the simple gauge boson legs with gravitons, and the $\binom{K+2}{3}$ equations of \cref{eqn:anomaly_u1_u1_u1} (one for each unordered combination of $k,k',k''\in\{1,\dots,K\}$) come from replacing them with abelian gauge boson legs.
		
	A subtlety, implicit in the discussion above, is that the data of a gauge quantum field theory that are relevant for anomalies include not just the Lie algebra and the fermion representation, but also a choice of ordered basis for the abelian summand. Indeed, the definition of a gauge quantum field theory includes specifying the kinetic term for the gauge fields, or equivalently specifying the values of $K$ gauge couplings, defined with respect to some chosen ordered basis. This extra data is relevant, albeit only weakly, for anomalies. One way to see that this must be so is to note that if any gauge coupling vanishes, there can be no corresponding anomaly.	One can also see it by noting that if it were not the case, the anomaly would be invariant under all automorphisms of $\mathfrak{g}$ that lift to the gauge Lie group. For the abelian summand, for example, these automorphisms form a group isomorphic to $GL(K,\Z)$, but these do not leave the anomaly invariant. The upshot of this for our purposes is that, strictly speaking, solutions $x_{ki}$ and  $x^\prime_{ki}$ should be considered equivalent if and only if they agree on the nose (which corresponds to the usual mathematical notion of equivalence of representations, i.e. up to inner automorphism). However, the anomaly turns out to also be invariant under the outer automorphisms of $\mathfrak{g}$ given by permuting the $K$ $\mathfrak{u}_1$ summands, so we will often consider solutions related in this way to be equivalent as well.
		
	As for the question of when a solution $x_{ki}$ should be regarded as physically acceptable, we shall insist that the $K$ quantities $x_{1i},x_{2i},\dots x_{Ki}$ are linearly independent for each $i$. For if they are not, we can change to a basis in which one or more of the $K$ gauge bosons does not couple to the fermions, and so cannot suffer from an anomaly.\footnote{A similar observation was made in the case $K=2$ in \cite{Costa_2020}, where the analogous condition is that $x_{1i},x_{2i}$ not be parallel. In the projective formalism that we shall soon introduce, they must be distinct projective points.}
		
	We shall say that a solution is \emph{chiral}, if the corresponding representation is not self-conjugate. Complex conjugation on irreducible representations sends $(x_{ki}, D_i^j ,T_i^j)$ to $(-x_{ki}, D_i^j ,T_i^j)$ and so a necessary condition for self-conjugacy is that a solution $x_{ki}$ corresponding to the $(I-1)$-fold reducible fermion representation be invariant under this action.\footnote{That is, the charges with the same values of $D_i^j ,T_i^j$ must either vanish or occur in opposite-sign pairs.}} This condition is obviously not quite sufficient for self-conjugacy because, as the action explicitly indicates, a representation of a simple Lie algebra and its conjugate have the same values of $D_i^j$ and $T_i^j$.\footnote{We remark that the condition of a representation being chiral is invariant under the full automorphism group $GL(K,\Z)$ of the abelian factor, since the action corresponds to the central element $-\mathbb{I} \in GL(K,\Z)$.}
	
	To describe the connection to algebraic geometry, begin by considering the case where there is only one $\mathfrak{u}_1$ summand; that is, when $K=1$. Since this case will play a dominant r\^{o}le in the sequel, we denote $x_{1i}$ by $x_i$ here for brevity.  
	
	\cref{eqn:anomaly_grav_grav_u1,eqn:anomaly_simple_simple_u1,eqn:anomaly_u1_u1_u1} now become
	\begin{align}
		\sum_{i=0}^{I-1} T_i^{j}\left(\prod_{\substack{j'=1 \\ j'\neq j}}^{J} D_i^{j'}\right) x_{i} &= 0,\,j\in\{1,\dots,J\}, \label{eqn:one_u1_anomaly_simple_simple_u1} \\
		\sum_{i=0}^{I-1}\left(\prod_{j=1}^{J} D_i^j\right) x_{i} &= 0, \label{eqn:one_u1_anomaly_grav_grav_u1}\\
		\sum_{i=0}^{I-1} \left(\prod_{j=1}^{J} D_i^j\right) x_{i}^3 &= 0. \label{eqn:one_u1_anomaly_u1_u1_u1}
	\end{align}
	
	These equations are all homogeneous in the $x_{i}$, which take values in the ring of integers, $\Z$, so we may just as well solve them over the field of rational numbers, $\Q$, discarding the trivial solution $(0,\dots,0)$. (Indeed, every integer solution is a rational solution, while we can turn a rational solution into an integer one by clearing denominators.) This is convenient, since we can then use constructions from geometry. In particular, we can define projective space as the set of equivalence classes $[x_0:\dots:x_{I-1}] \in \Q P^{I-1}$ (where two points in  $\Q^{I}$ are regarded as equivalent to another one if they differ by multiplication by a non-zero element in $\Q$) and the homogeneity of the above equations implies that we have a well-defined projective variety in $\Q P^{I-1}$. This involves no loss of information, since every affine solution can be recovered by multipling any representative of a projective solution by multiplying by all non-zero elements in  $\Q$. It is convenient because projective varieties are easier to work with in algebraic geometry than affine varieties, in much the same way that compact manifolds are easier to work with than non-compact ones.\footnote{It is also a convenient thing to do from a physics point-of-view, related to our discussion above of when two sets of charges should be considered equivalent. Namely, such a rescaling can be effected by rescaling the gauge coupling.} 
	
	Finally, it is convenient to assume that the linear equations here are linearly independent. Dependency can arise (for example if all fermions carry the same irreducible representation with respect to some simple factor, then the corresponding equation is a multiple of the equation corresponding to the gravitational contribution), but if it does one may simply discard some of the linear equations.
	
	Having done so, and assuming that $I-1>J$ (if not, things are much more straightforward), we can use the linear constraints to eliminate $J+1$ variables from \cref{eqn:one_u1_anomaly_u1_u1_u1} and conclude that the resulting homogeneous cubic equation in $I-J-1$ variables defines a projective cubic hypersurface in $\Q P^{I-J-2}$, which we call $X$.
	
	\subsection{Lines through points} \label{sec:failure_lines_through_point}
	
	At this point, we interrupt our general discussion to explain why the approach of \cite{Costa_2020}, which boils down to finding points on the cubic hypersurface and then points on lines (or more generally linear subspaces) through those points, does not work in general.
	In fact, there are at least four reasons why this approach fails. The first reason is that it requires us to be able to find enough points on the cubic hypersurface to be able to be sure of finding all linear subspaces. Clearly not all points are required to do this (since we only need to find one of the infinitely many points on each linear subspace), but it is unclear, in general, how one can figure out how many points are `enough’. The second reason is that no matter how many points are enough, one cannot be sure of finding them. Indeed, it is not known how to find even a single point on a cubic hypersurface, in general (assuming one exists).
	(The best one is able to do, as explained in \cite{Kollar_2002}, is to find infinitely many points given a single starting point.) For a third reason, suppose we have enough points. Given any such point, the second step is to try to construct all linear subspaces through it. A generic point is smooth, so let us assume that our given point is. By means of a projective transformation, we can without loss of generality assume that the point has homogeneous co-ordinates $[x_0:x_1:\dots:x_n]=[1:0:\dots:0]$ and that its tangent hyperplane is $x_1=0$. The equation defining the cubic hypersurface then takes the form
	\begin{gather}
		0 = x_0^2x_1 + x_0 Q(x_1,\dots x_n) +  x_0 C(x_1,\dots x_n),
	\end{gather}
	where $Q$ and $C$ are arbitrary quadratic and cubic homogeneous polynomials, respectively. A line in the hypersurface necessarily lies in the tangent plane $x_1=0$ and can be parameterized in terms of its direction $[v_2:\dots:v_n] \in \PP^{n-2}$ and a parameter $t$ along the line as $(x_2,\dots,x_n) = t(v_2,\dots v_n)$, and so lies in the hypersurface if and only the equations $Q(0,v_2,\dots v_n) = C(0,v_2,\dots v_n) = 0$ are satisfied.\footnote{The ground field is always taken to be $\Q$, so $\PP^n:=\Q P^n$, unless we explicitly say otherwise.} Now, if $n=3$ (which is the case of a cubic surface), these equations generally have no solution, while if $n \geq 4$ we are only guaranteed solutions over an algebraically-closed field. Working over $\Q$, as we must, there may be no solutions (an example is given by taking $Q = x_1^2+\dots+x_n^2$). The fourth reason is that even if solutions do exist, one needs to be able to find them. When $n=4$ this may be straightforward, since we just need to solve a quadratic in two unknowns and then factorize a cubic in a single unknown. But when $n>4$, this is again challenging.
	
	Since this particular strategy does not work, we thus return to formulating the general problem using algebraic geometry.
	The key idea is a simple one, namely that rather than looking for linear subspaces through a given point, one can look for linear subspaces directly. These linear subspaces assemble into varieties, called \emph{Fano varieties of linear subspaces}, which can themselves be studied using the methods of algebraic geometry.\footnote{The same is true for linear subspaces together with their points, which can be assembled into an \emph{incidence} variety, and used to study the `lines through points' approach using the tools of algebraic geometry.}
	We stress that this is not a panacea which allows us to solve all examples arising from physics. But it will allow us to solve a number of examples and will enable us to see more clearly where the difficulties in solving other examples lie.
	
	\subsection{Fano varieties of linear subspaces} \label{sec:fano_varieties}
	As we have described, solutions of the anomaly cancellation conditions with $K=1$ correspond to points of the projective variety $X$ (which we often call {\em rational points}). More precisely, one can recover all solutions of the anomaly cancellation conditions from them. We now wish to explain the crucial observation behind this paper, namely that the same is true for general $K$ when we replace `points of $X$' by `$K-1$-dimensional projective linear subspaces of $X$'.
	
	The proof of this fact is a triviality: a $(K-1)$-dimensional projective linear subspace of $\Q P^{I-1}$ can be parameterized as $[x_0:\dots:x_{I-1}]:=\sum_{k=1}^{K}\alpha_k[x_{k0}:\dots:x_{k,I-1}]$, where $\{[x_{k0}:\dots:x_{k,I-1}],1\leq k\leq K\}$ are $K$ points in $\Q P^{I-1}$ that are in general linear position and $[\alpha_1: \dots: \alpha_K] \in \Q P^{K-1}$ are the parameters. Plugging $[x_0:\dots:x_{I-1}]$ into \cref{eqn:one_u1_anomaly_grav_grav_u1,eqn:one_u1_anomaly_simple_simple_u1,eqn:one_u1_anomaly_u1_u1_u1} and insisting that they are solved for all values of $[\alpha_1, \dots, \alpha_K]$ reproduces \cref{eqn:anomaly_grav_grav_u1,eqn:anomaly_simple_simple_u1,eqn:anomaly_u1_u1_u1}. The argument also goes the other way, because of our earlier insistence that the $[x_{k0}:\dots:x_{k,I-1}]$ be linearly independent. 
	
	Again, we should be precise about what we mean by `correspond' here. Given a $(K-1)$-dimensional projective linear subspace in $X$, we can recover all solutions of the anomaly cancellation equations with $K$ $\mathfrak{u}_1$ summands as the ordered $K$-tuples of points of the subspace that are in general linear position. (So given a line in $X$, for example, we take any pair of distinct points on it, and any the solutions are given by any integer charges that land in the corresponding projective equivalence classes.)
	
	We shall say that a $(K-1)$-dimensional projective linear subspace in $X$ is \emph{chiral} if it contains a point corresponding to a chiral representation. We remark that for $K>1$ this does not imply that all points in the subspace are chiral,\footnote{A counterexample for $\mathfrak{u}_1\oplus\mathfrak{u}_1$ gauge theory is given in \cref{sec:u1u1}, where we exhibit chiral lines that intersect five non-chiral planes.} nor even that any set of $K$ points in general linear position contains a chiral point, so whether such solutions are chiral needs to be checked by hand. Happily, the chiral points form a non-empty open, ergo dense, subset, so this is rarely a problem. 
	
	At this juncture, we repeat that our reinterpretation of the anomaly cancellation conditions \cref{eqn:anomaly_grav_grav_u1,eqn:anomaly_simple_simple_u1,eqn:anomaly_u1_u1_u1} does not necessarily make them any easier to solve. As we will shall soon see, very little is known in general, even regarding the most coarse features of the set of solutions. However, it does at least allow us to make contact with much-studied objects in algebraic geometry and to carry over verbatim results that have been obtained in that field. As it turns out, that knowledge will enable us to solve the equations in certain specific examples. With this in mind, we recall some standard notions from algebraic geometry. 
	
	The \emph{grassmannian} $\mathbb{G}(k,n)$ is a variety assembled from the set of all $k$-dimensional projective linear subspaces (hereafter $k$-planes) of $n$-dimensional projective space $\PP^n$. Given a hypersurface $X\subset\PP^n$ of degree $d$, its $k$-planes assemble into a subvariety $F_k(X)\subset\mathbb{G}(k,n)$, the \emph{Fano variety of $k$-linear subspaces}. In the context of our anomaly cancellation problem, we thus need to understand $F_{K-1}(X)\subset\mathbb{G}(K-1,I-J-2)$ for a cubic hypersurface $X\subset\PP^{I-J-2}$.
	
	An obvious first question is: what is the dimension of $F_k(X)$? 
	Here, we already run into trouble, as the answer is not generally known. For a \emph{general} hypersurface over an algebraically-closed field, it can be shown that 
	\begin{equation}
		\dim F_k(X) = (k+1)(n-k)-\binom{k+d}{k},\label{eqn:dim_fano}
	\end{equation}
	where negative dimensions mean that $F_k(X)$ is empty \cite{Eisenbud_2016}. But the only result we have over $\Q$ is that the dimension of the Fano variety of lines on a smooth cubic hypersurface in $\PP^N$ is $2N-6$ \cite{Debarre_2016}. (Unfortunately the example we consider in \cref{sec:u1u1} is not smooth, but there our methods determine its dimension explicitly.)
	
	One can give a more explicit description of 
	the grassmannian (and hence the Fano variety embedded in it) in more than one way. We shall make use of two in studying examples from gauge theory. 
	In order to be able to get to the examples as soon as possible, we postpone discussion of these descriptions to \cref{app:fano}.
	
	\section{Rank two: cubic surfaces} \label{sec:cubic_surface}
	In this section, we consider examples from physics of
	gauge Lie algebras with abelian summand $\mathfrak{u}_1 \oplus \mathfrak{u}_1$ for which solving the anomaly cancellation conditions corresponds to finding lines on cubic surfaces. Famously, a smooth cubic surface over an algebraically-closed field has 27 lines, but we will see that in our context the story is much richer, though nevertheless tractable. 
	
	So, consider a theory with gauge Lie algebra $\mathfrak{su}_2\oplus\mathfrak{u}_1\oplus\mathfrak{u}_1$, where the six fermions transform in the representation $\bigoplus_{i=1}^6 d_i$ of the Lie algebra $\mathfrak{su}_2$ and the positive integer $d_i$ is the dimension of the $i$-th irreducible representation summand. As we have seen in \cref{sec:alggeo}, the possible anomaly-free charges correspond to lines in the cubic surface defined by 
	\begin{align}
		\sum_{i=0}^5 d_i(d_i^2-1)x_i&=0,\label{eqn:six_fermions_cubic}\\
		\sum_{i=0}^5 d_ix_i&=0,\label{eqn:six_fermions_linear_1}\\
		\sum_{i=0}^5 d_ix_i^3&=0,\label{eqn:six_fermions_linear_2}
	\end{align}
	where the $x_i$ correspond to possible charges of the fermions with a single $\mathfrak{u}_1$ factor.
	
	As explained in \cite{Gripaios_2026_u1}, different choices for the dimensions $d_i$ may lead to very different varieties, depending largely on the degeneracies of the values $d_i$, which defines a partition of 6. These include cases where the cubic surface is irreducible and either does or does not contain rational lines.\footnote{All choices of $d_i$ that we make in this Section are free of $SU(2)$ global anomalies in the sense of \cite{Witten_1982} and \cite{Wang_2019}. There is no pure gauge $\mathfrak{su}_2$ anomaly associated with any choice of $d_i$.}
	
	A smooth cubic surface over the complex numbers famously has 27 lines, but over the rationals
	Segre showed that the number of lines takes values in $\{0, 1, 2, 3, 5, 7, 9, 15, 27\}$ \cite{Segre_1949}.  
	For singular cubic surfaces, line counts over $\C$ are known and are determined by the number and types of singularities on the surface \cite{Bruce_1979,Sakamaki_2010}. The number of lines is one of $\{1,2,3,4,5,6,7,8,9,10,11,12,15,16,21,\infty\}$. As far as we know, there are no known results regarding what happens over $\Q$. 
	
	We will show how the Pl\"{u}cker embedding of the grassmannian described in \cref{app:fano}
	can be used to obtain the equations of any such lines, following an algorithm described in \cite{Boissiere_2007}
	and thus solve the anomaly cancellation conditions.
	
	For the partition $6=5+1$, for example, for which we set $d_0=d_1=d_2=d_3=d_4\neq d_5$, the linear constraints give $x_0+x_1+x_2+x_3+x_4=x_5=0$, and the cubic constraint then defines the Clebsch diagonal cubic surface,\footnote{The same equation also arises if we ask that the \emph{five} fermions with charges $(x_0,\dots,x_4)$ in a theory with gauge Lie algebra $\mathfrak{u}_1$ are anomaly-free \cite{Allanach_2020_geometricu1}, or if we ask for anomaly-free irreducible representations of a theory with simple gauge Lie algebra $\mathfrak{su}_5$ \cite{Gripaios_2024_irreducible}.}
	\begin{equation}
		\sum_{i=0}^4 x_i^3 = \sum_{i=0}^4 x_i = 0,
	\end{equation}
	or equivalently,
	\begin{equation}
		x_0(x_1+x_2+x_3)(x_0+x_1+x_2+x_3)+(x_1+x_2)(x_1+x_3)(x_2+x_3) = 0.
	\end{equation}
	We find the equations of the lines on the surface explicitly in \cref{app:clebsch_lines}. The Clebsch diagonal cubic surface contains 15 rational lines, all of which are non-chiral and of the form $x_i+x_j=x_k+x_l=x_m=0$ for $\{i,j,k,l,m\}=\{0,1,2,3,4\}$. It follows that there are no chiral set of six fermions that are anomaly-free under an $\mathfrak{su}_2\oplus\mathfrak{u}_1\oplus\mathfrak{u}_1$ theory where $d_0=d_1=d_2=d_3=d_4\neq d_5$.\footnote{As per the previous footnote, it follows that there are no anomaly-free chiral solutions for the charges of five fermions in a $\mathfrak{u}_1\oplus\mathfrak{u}_1$ gauge theory (a result previously obtained in \cite{Costa_2020} in a different way).} 
	
	To get chiral lines, consider the partition $6=3+3$, for which setting $d_0=d_1=d_4\neq d_2=d_3=d_5$ leads to the cubic surface defined by
	\begin{equation}
		d_0x_0x_1(x_0+x_1)+d_2x_2x_3(x_2+x_3)=0.
	\end{equation}
	Applying the algorithm shows that this contains nine non-chiral rational lines and six chiral \emph{real} lines of the form $[x_0:x_1:x_2:x_3]=[u:v:uc+vg:ud+vh]$ (thus all lying in the first Pl\"ucker stratum) for the combinations
	\begin{align*}
		c=h=0, d=g=-\lambda,&&  c=-d=g=\lambda, h=0,&& c=h=-\lambda, g=d=0, \\
		c=g=-h=\lambda, d=0,&&  c=-d=-h=-\lambda, g=0,&&	 c=0, d=-g=h=\lambda,
	\end{align*}
	where $\lambda:=\left(\frac{d_1}{d_3}\right)^{1/3}$. where $\lambda:=\left(\frac{d_1}{d_3}\right)^{1/3}$. Clearly these lines are rational lines iff $\lambda\in\Q$, i.e. if each exponent in the prime factorization of $\frac{d_1}{d_3}$ is divisible by 3.
	So we discover that the presence or absence of anomaly-free chiral fermions in this case is determined by the specific values of the dimensions of the irreducible representations of the simple summand. 
	
	To see just how rich things can be, let us consider some examples corresponding to the partition $6=1+1+1+1+1+1$. 
	
	On the one hand, the choice $(d_0,d_1,d_2,d_3,d_4,d_5)=(2,3,5,6,7,11)$ gives a cubic surface that with no rational line, which means that there
	are no anomaly-free gauge theories with fermions carrying these representations of the simple summand.
	
	On the other hand, the dimensions $(d_0,d_1,d_2,d_3,d_4,d_5)=(2,3,5,6,7,9)$ give rise to the cubic surface
	\begin{equation}
		2x_0^3+3x_1^3+5x_2^3+6x_3^3+\frac{(15x_0+20x_1+20x_2+13x_3)^3}{12\,288}-\frac{(77x_0+108x_1+140x_2+135x_3)^3}{200\,704}=0,
	\end{equation}
	on which we find three rational lines of the form ${[x_0:x_1:x_2:x_3]=[u:v:uc+vg:ud+vh]}$, all of which are in the first Pl\"ucker stratum and are, of necessity, chiral:
	\begin{align*}
		c=\frac{102}{53}, d=-\frac{63}{53}, h=-\frac{464}{265}, g=\frac{567}{265}; 
		&& c=\frac{60}{109}, d=-\frac{63}{109}, h=-\frac{184}{545}, g=-\frac{567}{545}; \\
		c=-\frac{3}{4}, d=0, h=-\frac{4}{5}, g=0.	
	\end{align*}
	This last example is somewhat amusing since, while we can (indeed we just have) find all solutions to the anomaly cancellation equations for the $\mathfrak{su}_2 \oplus \mathfrak{u}_1 \oplus \mathfrak{u}_1$, finding all solutions to the corresponding $\mathfrak{su}_2 \oplus \mathfrak{u}_1$ case is currently beyond us. This can be traced to the fact that the three lines we have found intersect pairwise, so can not be used to show that the surface is a rational variety. The best one can do is to show that it is a unirational variety; doing so enables us to find infinitely many solutions, but certainly not all of them. 
	\section{Rank two: the Segre cubic primal 3-fold} \label{sec:u1u1}
	When we increase the number of fermions or decrease the number of simple summands, the dimension of the cubic hypersurface goes up and so does the (expected) dimension of the Fano variety of linear subspaces. This allows life to get a lot more interesting, as we will now see for the simplest non-trivial case of pure $\mathfrak{u}_1 \oplus \mathfrak{u}_1 $ gauge theory with six fermions.\footnote{The same cubic hypersurface is obtained for the $\mathfrak{su}_2\oplus \mathfrak{u}_1 \oplus \mathfrak{u}_1$ gauge theories considered in the last section when all six $d_i$'s are equal, because the two linear constraints \cref{eqn:six_fermions_linear_1,eqn:six_fermions_linear_2} are the same.}
	
	In this case, the anomaly cancellation equations for the single $\mathfrak{u}_1 $ case become
	\begin{equation}
		x_0^3+x_1^3+x_2^3+x_3^3+x_4^3+x_5^3=x_0+x_1+x_2+x_3+x_4+x_5=0, \label{eqn:segres6}
	\end{equation}
	or equivalently,
	\begin{equation}
		x_0^3+x_1^3+x_2^3+x_3^3+x_4^3-(x_0+x_1+x_2+x_3+x_4)^3=0,\label{eqn:segre}
	\end{equation}
	with
	\begin{equation}
		x_5=-(x_0+x_1+x_2+x_3+x_4).\label{eqn:x5}
	\end{equation}
	\cref{eqn:segre} defines a famous cubic \emph{3-fold} (\emph{i.e.} a three-dimensional cubic hypersurface) ${\mathfrak{S}\subset\PP^4}$ called the \emph{Segre cubic primal}. Like its lower-dimensional cousin the Clebsch diagonal cubic surface, it has been an object of intense study in classical and modern algebraic geometry for more than one and a half centuries \cite{Segre_1886,Castelnuovo_1891,Richmond_1902,Manin_1986,Dolgachev_2012,Dolgachev_2016,Ciliberto_2024_gromov,Ciliberto_2024_modular}, and it has appeared in very recent theoretical physics literature in the context of positive geometries, scattering amplitudes and the amplituhedron \cite{Early_2017,Tevelev_2025,Sturmfels_2026}. Unlike the Clebsch diagonal cubic surface, it is singular, containing 10 double points whose coordinates are obtained as permutations under the $S_6$ symmetry manifest in \cref{eqn:segres6} of the point ${[x_0:x_1:x_2:x_3:x_4:x_5]=[1:1:1:-1:-1:-1]}$ and 15 non-chiral planes $(ij|kl|mn)$ obtained as permutations of the plane with equation ${x_0+x_1=x_2+x_3+x_4+x_5=0}$, which we denote $(01|23|45)$. Every double point lies on six planes: for example, the double point $[1:1:1:-1:-1:-1]$ lies on the planes $(03|14|25), (03|15|24), (04|13|25), (04|15|23), (05|13|24)$ and $(05|14|23)$. 
	
	The existence of double points means that the Segre cubic primal is a rational cubic threefold; that is, it is birationally equivalent to $\Q P^3$. An explicit birational map can be got from projecting from any one of the double points, as described in \cite{Allanach_2020_geometricu1}.
	
	Let us now investigate the lines on the Segre cubic primal, which will yield solutions to the anomaly cancellation equations of a $\mathfrak{u}_1\oplus\mathfrak{u}_1$ gauge theory with six fermions.
	
	We will use two distinct approaches. The first approach has the advantage of being applicable (at least a priori) to other examples (indeed, we will use it to study the $\mathfrak{u}_1\oplus\mathfrak{u}_1$ case with one more fermion in \cref{sec:otherex} and the $\mathfrak{u}_1\oplus\mathfrak{u}_1\oplus\mathfrak{u}_1$ case elsewhere \cite{Gripaios_unpublished}), and allows us to show that the Fano variety of lines on $\mathfrak{S}$ has 21 irreducible components, all of dimension 2, 15 of which are planes corresponding to (non-chiral) lines lying in the 15 planes of $\mathfrak{S}$, while the other 6 components consist of chiral lines, are all isomorphic to one another, and are both smooth and rational.
	
	From here, one could go on to determine the isomorphism class of the chiral components as that of the split del Pezzo surface of degree 5 directly. But rather than reinvent the wheel, it is easier to follow the existing second approach. This also has the advantage of being geometrically more insightful. Its disadvantage is that it is specific to this example.
	
	\subsection{First approach: an affine cover of the grassmannian} \label{sec:segre_cover}
	In the first approach, we begin by defining a cover of the grassmannian $\mathbb{G}(1,4)$ of projective lines in $\PP^4$. As described in \cref{app:fano}, the lines not intersecting any chosen plane define an affine open set. We choose not just any old plane in $\PP^4$ but rather the plane $(01|23|45)$ lying in $\mathfrak{S} \subset\PP^4$, for two reasons. Firstly, it turns out that for such planes, the polynomials defining points in $F_1(\mathfrak{S})$ are not cubic but (at most) quadratic, making them much easier to solve. Indeed, we will see that it is easy to identify the irreducible components of the variety $F_1(\mathfrak{S})$ that intersect this affine open set and to show that they are rational varieties. Secondly, as we show in \cref{app:cover}, the affine open sets defined by all 15 planes in $\mathfrak{S}$ cover $\mathbb{G}(1,4)$ and hence cover $F_1(\mathfrak{S})$. Thus we can study $F_1(\mathfrak{S})$ globally in this way. A word of warning: since the grassmannian has dimension $(1+1)[(4+1)-(1+1)] = 6$ and since each component of the Fano variety will turn out to be two-dimensional, it is quite possible (and indeed it turns out to be the case) that $F_1(\mathfrak{S})$ has components that have empty intersection with one of these open sets. So a global study is necessary.
	
	Thus, let us start by considering the open subset $U_{01,23,45}$ of $F_1(\mathfrak{S})$ containing lines on $\mathfrak{S}$ that miss the plane $(01|23|45)$. Defining new coordinates ${y_0:=\frac{x_0+x_1}{2}}$, ${y_1:=\frac{x_2+x_3}{2}}$, ${y_2:=\frac{x_0-x_1}{2}}$, ${y_3:=\frac{x_2-x_3}{2}}$, the plane $(01|23|45)$ is given by the vanishing of $y_0$ and $y_1$, and so any line on $\mathfrak{S}$ missing this plane is given by the row space of the matrix $\begin{pmatrix}	1 & 0 & a_2 & a_3 & a_4 \\ 0 & 1 & b_2 & b_3 & b_4 \end{pmatrix}$. In the old coordinates $x_i$, the same line is given by the row space of $\begin{pmatrix} 1+a_2 & 1-a_2 & a_3 & -a_3 & a_4 \\ b_2 & -b_2 & 1+b_3 & 1-b_3 & b_4 \end{pmatrix}$; that is, the coordinates of a point $[x_0:x_1:x_2:x_3:x_4]$ on this line are parameterized by
	\begin{equation}
		[(1+a_2)t+b_2u:(1-a_2)t-b_2u:a_3t+(1+b_3)u:-a_3t+(1-b_3)u:a_4t+b_4u],
	\end{equation}
	where $[t:u]\in\Q P^1$ is a parameter along the line. Such a line lies on $\mathfrak{S}$ if
	\begin{multline}
		a_2^2-(1+a_4)^2=a_3^2+2a_2b_2-2b_4(a_4+1)-(a_4+2)^2=\\
		=b_2^2+2a_3b_3-2a_4(b_4+1)-(b_4+2)^2=b_3^2-(b_4+1)^2=0. \label{eqn:u012345}
	\end{multline}
	
	Notice that, as claimed above, these equations are all quadratic in the $a_i$'s and $b_i$'s. Solving these equations gives us two distinct types of solutions.
	
	The first type of solutions are linear 2-dimensional subspaces of $F_1(\mathfrak{S})$, of which there are eight with $(a_3,a_4,b_3,b_4)$ given below,
	\begin{align*}
		(-1-a_2,-1-a_2,-1-b_2,-2-b_2), && (1-a_2,-1-a_2,1-b_2,-b_2), \\
		(-1-a_2,-1+a_2,-1-b_2,b_2), && (1-a_2,-1+a_2,1-b_2,-2+b_2), \\
		(-1+a_2,-1-a_2,-1+b_2,-b_2), && (1+a_2,-1-a_2,1+b_2,-2-b_2), \\
		(-1+a_2,-1+a_2,-1+b_2,-2+b_2), && (1+a_2,-1+a_2,1+b_2,b_2).
	\end{align*}
	These evidently correspond to the lines lying on the (non-chiral) planes $(ij|kl|mn)$ on $\mathfrak{S}$ that, by our choice of affine open set, do not also lie on the plane $(01|23|45)$. Here we see just 8 of the 15 planes,\footnote{In the order listed above, they are $(02|15|34)$, $(04|13|25)$, $(02|14|35)$, $(05|13|24)$, $(04|12|35)$, $(03|15|24)$, $(05|12|34)$, $(03|14|25)$.} corresponding to the fact that for 7 of the planes, every line in them intersects the chosen plane $(01|23|45)$. One of these 7 is of course the plane $(01|23|45)$ itself, and, as we shall later explicitly show, there are 6 others which intersect this plane in a line. This confirms that there are indeed irreducible components of $F_1(\mathfrak{S})$ that do not intersect this affine open set, and so we must consider further affine open sets in order to find all lines in $\mathfrak{S}$.
	
	Before doing so, let us consider the remaining solutions to \cref{eqn:u012345}. These define four, irreducible, two-dimensional components, given by the equations
	\begin{align}
		D_0&: a_2=-a_4-1, b_3=-b_4-1, a_2+a_3=b_2+b_3, (a_2+a_3)^2=4a_2b_3+1.\\
		D_3&: a_2=-a_4-1, b_3=b_4+1, a_2-a_3=b_2-b_3, (a_2-a_3)^2=-4a_2b_3+1. \\
		D_4&: a_2=a_4+1, b_3=b_4+1, a_2+a_3=b_2+b_3, (a_2+a_3)^2=4a_2b_3+1.  \\
		D_5&: a_2=a_4+1, b_3=-b_4-1, a_2-a_3=b_2-b_3, (a_2-a_3)^2=-4a_2b_3+1. 
	\end{align}
	(The apparently peculiar labelling is chosen to agree with the conventional choices in the second approach, described in the next Subsection.)

A first observation is that each irreducible component is a smooth variety. Indeed, none of the four components has a singular point in this affine open set and this implies that no irreducible component has a singular point in any of the affine open sets in the cover, because each can be reached from any other by permutation.

A second observation is that each component is unirational, for we
can use $(a_2,a_3)$ as coordinates on $\mathbb{A}^2$ (which is birationally equivalent to $\PP^2$) to
define a dominant rational map in each case.
Since this map is generically two-to-one, it cannot be a birational map. So to show that our varieties are rational, we will need to find another map. This is easily done, for each of these components is determined by linear equations along with a single quadratic equation in 3 unknowns, which defines a quadric surface in 3-dimensions. Moreover, one checks that for each component,  this quadric surface has a rational point, so a birational map from $\PP^2$ can be defined by finding the other intersection of each line emanating from such a point with the surface.\footnote{Perhaps more elegantly, we can use the fact that each quadric surface is smooth and admits a rational line, so is therefore a doubly-ruled surface, whose two families of lines can be used to give a birational map. The component $D_1$, for example, becomes $(a_2+a_3)^2 = 4a_2 b_3 + w^2 \implies (a_2+a_3+w)(a_2+a_3-w) - 2a_2.2b_3 = 0$ after homogenizing, so we can parametrize points on it as $a_2 = \frac{ut}{us-vt}, a_3 = \frac{us + vt - ut}{us-vt}, b_3 = \frac{vs}{us-vt}, w = 1$ in terms of co-ordinates $([s:t],[u:v]) \in \PP^1 \times \PP^1$.}

Finally, we observe that all these 4 components are isomorphic; this follows since there exist permutations of the $x_i$ sending them into one another. 

What about further irreducible components that do not intersect this affine open set? We already know that there are 7 planes missing. Moreover, we know that if there are any further irreducible components, they must be isomorphic to the components $D_i$, because we know that $S_6$ acts transitively on the set of 15 planes in $\mathfrak{S}$ ergo transitively on the set of 15 affine open sets making up our cover of $F_1(\mathfrak{S})$. The same logic shows that each affine open set intersects 4 $D$ components. The question then arises of how many of these 60 $D$-component avatars are independent components. The answer turns out to be 6. This can be shown exhaustively by considering each affine open set in turn and determining whether its 4 $D$ components intersect any of the previous ones in an open set. Alternatively, it can be shown by observing that the $S_6$ symmetry must acts transitively on the set of $D$ components, while each component is stabilized by a subgroup isomorphic to $S_5$ (as can be shown by acting with $S_6$ on a suitably small non-empty open set of any one component in this patch and checking whether or not it is contained in the same component in this patch). Hence the size of the orbit is $6!/5! = 6$.

Our methods so far already enable us to parametrize almost all (i.e. those in a dense open set) rational lines on the surface, corresponding to chiral solutions of the anomaly cancellation conditions. But of course, we would like to go further and be sure of being able to find all solutions. We would also like to better understand the $D$ components. To do so, we follow in the footsteps of the algebraic geometers of old, making use of a different rational map.

\subsection{Second approach: birational equivalences}
A large amount of work has already gone into understanding $F_1(\mathfrak{S})$ (at least over $\mathbb{C}$), in particular \cite{Castelnuovo_1891,Richmond_1902,Dolgachev_2016}. These proceed by considering the four-dimensional\footnote{A projective quadric in $\PP^3$ is defined by the equation $\sum_{0\leq i\leq j\leq 3}a_{ij}x_i x_j=0$ for $a_{ij}\in\K$. There are $\binom{3+2}{2}=10$ monomials in this expression, so the space of quadrics in $\PP^3$ is isomorphic to $\PP^9$. Requiring these quadrics to pass through five points in general position then brings the space of these quadrics down to $\PP^4$.} linear system of quadrics that pass through five points in general position in $\PP^3$ and show that the image of the rational map $\PP^3\dashrightarrow\PP^4$ determined by these quadrics\footnote{As usual, the broken arrow $\dashrightarrow$ indicates that the map is not necessarily defined on the whole of $\PP^3$; indeed, here it is undefined at the five chosen points.} is the Segre cubic primal $\mathfrak{S}$. 

To be explicit, consider the rational map $f:\PP^3\dashrightarrow\mathfrak{S}$ given in \cite{Richmond_1902} by
\begin{equation}
	f:\PP^3\ni[a:b:c:d]\mapsto\begin{bmatrix}
		x_0 \\ x_1 \\ x_2 \\ x_3 \\ x_4 \\ x_5
	\end{bmatrix}
	:=
	\begin{bmatrix}
		-ac+bc+ad-bd-ab+cd \\
		-ac+bc+ad-bd+ab-cd \\
		+ac+bc-ad+bd-ab-cd \\
		-ac-bc-ad+bd+ab+cd \\
		+ac-bc+ad+bd-ab-cd \\
		+ac-bc-ad-bd+ab+cd
	\end{bmatrix}
	\in\mathfrak{S}.\label{eqn:f}
\end{equation}
\begin{sloppypar}
	We have written this as a map to $\PP^5$ rather than $\PP^4$ so as to make the symmetry with respect to permutations of the $x_i$ as manifest as possible, but it is evident that the map lands in the subspace $\PP^4 \subset \PP^5$ defined by $\sum_{i=0}^6x_i = 0$. It corresponds to the linear system of quadrics through the five points $(p_1,p_2,p_3,p_4,p_5)=({[1:0:0:0]},{[0:1:0:0]},{[0:0:1:0]},{[0:0:0:1]},{[1:1:1:1]})$ in $\PP^3$. Evidently the map is undefined at those points, since it sends them to $[0:0:0:0:0:0]\notin \PP^5$, but is defined at all other points, since one easily checks that 
	points in the image satisfy \cref{eqn:segre}. 
\end{sloppypar}

To show that $f$ is a birational equivalence, we exhibit an inverse. Namely, consider the map $g:\mathfrak{S}\dashrightarrow\PP^3$ given by
\begin{equation}
	g:\mathfrak{S}\ni
	\begin{bmatrix}
		x_0 \\ x_1 \\ x_2 \\ x_3 \\ x_4 \\ x_5
	\end{bmatrix}
	\mapsto
	\begin{bmatrix}
		a \\ b \\ c \\ d
	\end{bmatrix}
	:=
	\begin{bmatrix}
		+(x_0+x_4)(x_1+x_3)(x_2+x_5) \\
		-(x_0+x_4)(x_1+x_3)(x_3+x_4) \\
		-(x_1+x_2)(x_1+x_3)(x_2+x_5) \\
		-(x_0+x_4)(x_0+x_5)(x_2+x_5)
	\end{bmatrix}
	\in\PP^3, \label{eqn:g}
\end{equation}
where again $x_5$ is given by \cref{eqn:x5}. This map is defined everywhere on $\PP^4 \subset \PP^5$ apart from the four planes (which lie on $\mathfrak{S}$) given by $(04|12|35),(04|13|25),(05|13|24),(01|25|34)$, so it is defined everywhere on  $\mathfrak{S} \subset \PP^4$ away from those planes. Thus $g$ is a rational map. We show in \cref{app:inverse} that it inverts $f$.

For later purposes, it is useful to consider which subvarieties of the domain of $f$ get contracted to points, i.e. to specify where the map fails to be injective. On such subvarieties, the inverse map must be undefined, so $f$ must send such subvarieties to one of the 4 planes just described. The preimages of these planes are themselves unions of planes in $\PP^3$ through the five special points $[1:0:0:0],[0:1:0:0],[0:0:1:0],[0:0:0:1],[1:1:1:1]$ and by an explicit computation one can show that the only subvarieties of $\PP^3$ that are contracted to points by $f$ are the ten lines $\ell_{\alpha \beta}$ ($1\leq \alpha < \beta\leq 5$) passing through pairs $p_\alpha$ and $p_\beta$ of these points. They are sent to double points, as follows:
\begin{align}
	\ell_{12}\mapsto[1:-1:1:-1:1:-1],&&\ell_{13}\mapsto[1:1:-1:1:-1:-1], \nonumber \\
	\ell_{14}\mapsto[1:1:-1:-1:1:-1],&&\ell_{15}\mapsto[1:-1:1:1:-1:-1], \nonumber \\
	\ell_{23}\mapsto[1:1:1:-1:-1:-1],&&\ell_{24}\mapsto[1:1:-1:-1:-1:1], \nonumber \\
	\ell_{25}\mapsto[1:-1:-1:-1:1:1],&&\ell_{34}\mapsto[1:-1:-1:1:-1:1], \nonumber \\
	\ell_{35}\mapsto[1:-1:1:-1:-1:1],&&\ell_{45}\mapsto[1:-1:-1:1:1:-1].
\end{align}
\subsubsection{The induced map on lines} \label{sec:induced_maps}
We are now ready to study the lines in $\mathfrak{S}$, which, we remind the reader, correspond physically to solutions to the anomaly cancellation conditions.

Given that Richmond's map $f$ is birational, we can understand the lines in $\mathfrak{S}$ (or at the very least those lines that intersect an affine open set of $\mathfrak{S}$) if we can understand their preimages in $\PP^3$.

As it turns out, many of the lines in  $\mathfrak{S}$ have preimages that are themselves lines in $\PP^3$, so let us begin by discussing these.

We have just seen that the 10 lines 
$\ell_{\alpha \beta}$ get sent to (double) points of $\mathfrak{S}$, so these are no good.
But there are two closely-related classes of lines in $\PP^3$ that do get sent to lines. 
Firstly, we have the other lines lying in the ten planes $[\alpha \beta \gamma]$ on $\PP^3$ passing through $p_\alpha$, $p_\beta$ and $p_\gamma$ ($1\leq \alpha<\beta<\gamma \leq 5$); these 10 planes
are mapped to ten planes on $\mathfrak{S}$ (and the other lines in them are mapped to lines) in the following way:
\begin{align}
	[123]\mapsto(05|14|23),&&[124]\mapsto(03|12|45),&&[125]\mapsto(01|25|34), \nonumber \\
	[134]\mapsto(02|15|34),&&[135]\mapsto(04|12|35),&&[145]\mapsto(05|13|24), \nonumber \\
	[234]\mapsto(04|13|25),&&[235]\mapsto(03|15|24),&&[245]\mapsto(02|14|35),&&[345]\mapsto(01|23|45). \label{eqn:planes1}
\end{align}

The other class of lines in $\PP^3$ that get sent to lines consist of 
lines intersecting just one of the five points $p_\alpha$. For each point, there is a $\PP^2$'s worth of such lines (corresponding to the different directions in which they may emanate). These do not get mapped into the known planes in $\mathfrak{S}$, and therefore must get sent to five $D$ components; we will label the component whose preimage is the $\PP^2$ of lines through $p_\alpha \in\PP^3$ by $D_\alpha$.\footnote{The labelling agrees with the labelling already given to the non-linear components in the affine open set $U_{02,13,45}$ in the previous subsection.} 

In \cref{app:dP}, we check this explicitly, showing that three of these components intersect $U_{01,23,45}$, while the remaining two intersect $U_{02,13,45}$ (one component intersects both).

The results of the previous subsection indicate that we still have 5 planes and 1 $D$ component to find (the latter must, moreover, intersect $U_{01,23,45}$). These components do not have preimages that are lines in $\PP^2$.

As it turns out \cite{Dolgachev_2016}, the preimage of the remaining $D$ component, which we label by $D_0$, is the two-dimensional space of twisted cubics passing through all five points $p_1,\dots,p_5$ in $\PP^3$. The standard \emph{twisted cubic} is the image of the projective line under the degree 3 \emph{Veronese embedding} $v:\PP^1\to\PP^3$ given by $[s:t]\mapsto[s^3:s^2t:st^2:t^3]$ and any twisted cubic is obtained from it by an automorphism of $\PP^3$, giving a space of dimension equal to that of $PGL(4)$ minus that of $PGL(2)$, namely 12; fixing 5 projective points in $\PP^3$ thus leaves a space of dimension 2. Explicitly, we can describe an affine open set in the space of twisted cubics through $p_1,\dots,p_5$ by means of \begin{gather} \label{eq:twist}
	[s:t]\mapsto [(s-t)(s-vt)(s-ut):s(s-t)(s-vt):s(s-t)(s-ut):s(s-ut)(s-vt)],
\end{gather}
with affine parameters $(u,v) \in \mathbb{A}^2$.
Plugging into $f$, we get that any such twisted cubic curve is sent to the following line,
\begin{equation}
	f:
	\begin{bmatrix}
		(s-t)(s-vt)(s-ut) \\
		s(s-t)(s-vt) \\
		s(s-t)(s-ut) \\
		s(s-ut)(s-vt)
	\end{bmatrix}
	\mapsto
	\begin{bmatrix}
		1-u+v \\
		-1+u-v \\
		-1+u+v \\
		1+u-v \\
		1-u-v \\
		-1-u+v
	\end{bmatrix}s+
	\begin{bmatrix}
		-u-v+uv \\
		-u+v+uv \\
		u-v-uv \\
		-u+v-uv \\
		u-v+uv \\
		u+v-uv
	\end{bmatrix}t. \label{eqn:D0}
\end{equation} 
Such a line evidently intersects the planes $(01|24|35)$ and $(05|12|34)$ when $s=0$ and $t=0$ respectively; we will later see that it intersects three more, and only three more, planes which are $(03|14|25)$, $(04|15|23)$ and $(02|13|45)$. It does not, however, lie entirely on any plane on the Segre cubic primal.


\sloppy{The remaining five planes are not in the image of $f$, but they can be recovered by blowing up $\PP^3$ at the five special points $p_i$ where $f$ is undefined \cite{Dolgachev_2016}. Let us here show this explicitly. We recall that the \emph{blow-up} of $\PP^n$ at the point ${P=[1:0:0:\dots:0]}$, for example, is the projective subvariety $B_1 \subset\PP^n\times\PP^{n-1}$ containing pairs ${([z_0:z_1:\dots:z_n],[y_1:\dots:y_n])\in\PP^n\times\PP^{n-1}}$ satisfying ${z_iy_j-z_jy_i=0\,\forall\,i,j\in\{1,\dots,n\}}$.\footnote{In simple terms, the effect of blowing up a projective plane at a point, for example, is to insert a fibre over that point parametrizing the tangent directions there.} By the universal property, we then get a morphism $\tilde{f}$ from the blow-up $B_5$ in five points to $\mathfrak{S}$ which agrees with the composition of the blow-down map $B_5 \to \PP^3$ with $f$ wherever this is defined. The claim is that each of the planes in $B_5$ given by the exceptional divisor $E_\alpha$ at $p_\alpha$ gets sent to one of the 5 remaining planes  in $\mathfrak{S}$ (and more precisely, lines in planes get sent to lines in planes).}

To show this, consider for instance the point $p_1 = [1:0:0:0] \in \PP^3$ and choose the affine patch on $B_5$ given by $y_1=1$, such that $z_2 = y_2 z_1, z_3 = y_3 z_1$. Plugging into $f$ allows us to determine $\tilde{f}$ on this patch; we get
\begin{equation}
	f:[1:z_1:y_2z_1:y_3z_1]\mapsto
	\begin{bmatrix}
		-y_2+z_1y_2+y_3-z_1y_3-1+z_1y_2y_3 \\
		-y_2+z_1y_2+y_3-z_1y_3+1-z_1y_2y_3 \\
		+y_2+z_1y_2-y_3+z_1y_3-1-z_1y_2y_3 \\
		-y_2-z_1y_2-y_3+z_1y_3+1+z_1y_2y_3 \\
		+y_2-z_1y_2+y_3+z_1y_3-1-z_1y_2y_3
	\end{bmatrix}
	\in\mathfrak{S}.\label{eqn:tildef}
\end{equation}
and, because we have removed a factor of $x_1$ on the right-hand side, we get an expression defined everywhere on the patch (including at $x_1=0$), which defines $\tilde{f}$. In particular,
restricting to the exceptional divisor, we get
\begin{equation}
	f:[1:z_1:y_2z_1:y_3z_1]\mapsto
	\begin{bmatrix}
		-y_2+y_3-1 \\
		-y_2+y_3+1 \\
		+y_2-y_3-1 \\
		-y_2-y_3+1 \\
		+y_2+y_3-1 \\
		+y_2-y_3+1
	\end{bmatrix}
	\in\mathfrak{S}.\label{eqn:tildef}
\end{equation}

This is evidently a plane; in fact, it is the plane $(05|12|34)$. By doing the same for the other four exceptional divisors, we obtain the following images of the five $E_\alpha$ on $\mathfrak{S}$ under $\tilde{f}$:
\begin{align}
	E_1\mapsto(05|12|34),&&E_2\mapsto(03|14|25),&&E_3\mapsto(04|15|23),&&E_4\mapsto(02|13|45),&&E_5\mapsto(01|24|35).\label{eqn:planes2}
\end{align}
With these 5 planes, we have accounted for all fifteen planes on $\mathfrak{S}$. We remind the reader that they are of less interest to us than the $D$ components, since they correspond to non-chiral solutions. 

We now describe how the isomorphism class of the $D$ components, which correspond to chiral solutions, can be determined. As well as being intellectually satisfying, this will provide us with an explicit isomorphism to a single $D$ component that can be used to parametrize all solutions in that component (rather than a dense set of solutions, which is what a birational equivalence achieves). By permutation, one can then find all solutions in all other $D$ components.

It will turn out that a $D$ component is isomorphic to a del Pezzo surface of degree 5, which we define to be the projective plane blown up in 4 points (in general position). This has already been shown in \cite{Dolgachev_2016}, implicitly over an algebraically-closed field. Over the rationals, we must be a little more careful, not least because there are many distinct isomorphism classes of del Pezzo surfaces of degree 5 over $\Q$ \cite{Skorobogatov_2001}. Fortunately, ours is uniquely identified by the fact (as we have already seen) that its automorphism group is $S_5$. For the reader's benefit, we will try to get away with an approach that involves as little knowledge 
of algebraic geometry as possible. Our strategy is as follows. We already described (in words) above how Richmond's map $f$ induces a birational\footnote{Indeed, $g_*$ defines a rational inverse.} map $f_*$ from $\PP^2$, parametrizing the lines through (say) the point $p_1 = [1:0:0:0] \in \PP^3$ to $D$. We will show that this map is defined everywhere except at the four points corresponding to lines that also pass through one of the other four special points $p_2,\dots,p_5 \in \PP^3$. By blowing up $\PP^2$ at these four points, we obtain an explicit birational morphism $\tilde{f_*}: B_4 \to D$. We will show that this morphism is finite by showing that it does not contract any of the ten ($-1$)-curves in $B_4$ to a point. Since we have already shown that $D$ is smooth, \emph{ergo} normal, it follows by Zariski's Main Theorem that our birational morphism is an isomorphism. This isomorphism then can be used to explicitly parametrize all solutions to the anomaly cancellation conditions lying on a single $D$ component in one fell swoop, with all solutions on other $D$ components obtained simply by permutation of the charges. 

Proceeding, write a line through $p_1 = [1:0:0:0] \in \PP^3$ as $[\alpha:\beta z_1: \beta z_2:\beta z_3]$, where $[ z_1: z_2: z_3] \in \PP^2$ fixes the direction and $[\alpha:\beta ] \in \PP^1$ is a parameter along the line. Plugging this into \cref{eqn:f} for $f$, we get the induced map
\begin{equation}
	f_*:\PP^2\ni[z_1:z_2:z_3]\mapsto\begin{bmatrix}
		x_0 \\ x_1 \\ x_2 \\ x_3 \\ x_4 \\x_5
	\end{bmatrix}
	:=
	\begin{bmatrix}
		-\alpha z_2+\beta z_1 z_2+\alpha z_3-\beta z_1 z_3-\alpha z_1+\beta z_2 z_3 \\
		-\alpha z_2+\beta z_1 z_2+\alpha z_3-\beta z_1 z_3+\alpha z_1-\beta z_2 z_3 \\
		+\alpha z_2+\beta z_1 z_2-\alpha z_3+\beta z_1 z_3-\alpha z_1-\beta z_2 z_3 \\
		-\alpha z_2-\beta z_1 z_2-\alpha z_3+\beta z_1 z_3+\alpha z_1+\beta z_2 z_3 \\
		+\alpha z_2-\beta z_1 z_2+\alpha z_3+\beta z_1 z_3-\alpha z_1-\beta z_2 z_3 \\
		+\alpha z_2-\beta z_1 z_2-\alpha z_3-\beta z_1 z_3+\alpha z_1+\beta z_2 z_3
	\end{bmatrix}
	\in D,\label{eqn:fstar}
\end{equation}
where the points in $D$ are defined implicitly by the lines in $\mathfrak{S}$, albeit in a less-than-pleasant way.
One easily shows that this map is undefined precisely at the points $[ z_1: z_2: z_3] \in \{[1 :0 :0 ], [0 :1 :0 ], [0:0 :1 ], [1 :1 :1 ]\}$. (For example, at $[1:0:0]$ the right-hand side becomes $[-1:+1:-1:+1:-1:+1 ]$, which is not a line, but a point.) Blowing up at these four points, we obtain a morphism $\tilde{f_*}:B_4 \to D$. To show that it is finite, we must show that it doesn't contract any of the 10 $(-1)$-curves in $B_4$ to a point. These 10 curves are given by the 4 exceptional divisors and the six strict proper transforms of the lines through two of the 4 points.

For the latter, our work is easily done: by continuity the map $\tilde{f_*}$ agrees with the composition of the blow-up map $B_4 \to \PP^2$ and $f: \PP^2 \dashrightarrow D$ when $f$ is defined, that is away from the blow up points. So for the $(-1)$-curve corresponding to the line $[z_1 : z_2 :0 ]$ through $[1 :0 :0 ]$ and $[0 :1 :0 ]$, it suffices to show that there are two points on the line, excluding $[1 :0 :0 ]$ and $[0 :1 :0 ]$ (where it is not defined and which are excluded from the strict proper transform), with distinct images under $f$ in $D$. For example, the line with $[z_1 : z_2 :0 ] = [1 : 1 :0 ]$ goes through the point $[-1 : 0 :0:0:0:1 ] $ in $\mathfrak{S}$ at parameter value $[\alpha:\beta ] =  [1:0 ] $, but this point is not on the line with $[z_1 : z_2 :0 ] = [1 : -1 :0 ]$. So the two lines cannot be the same and $\tilde{f}$ cannot contract the proper strict transform of the line $[z_1 : z_2 :0 ] \in \PP^2$ to a point.

For the 4 exceptional divisors, we must do a smidgeon more work, in that we must actually construct the blow-up. Consider, for example, the divisor over $[1 :0 :0 ]$. An affine patch that intersects this is given by $(s,t) \in \mathbb{A}^2$ on which we can figure out the action of $\tilde{f_*}$ by the ruse of evaluating $f_*$ on $[1:s:st]$. We get
\begin{equation}
	\tilde{f_*}(s,t)
	=
	\begin{bmatrix}
		-\alpha +\beta +\alpha t-\beta t-\alpha +\beta st \\
		-\alpha +\beta +\alpha t-\beta t+\alpha -\beta st \\
		+\alpha +\beta -\alpha t+\beta t-\alpha -\beta st \\
		-\alpha -\beta -\alpha t+\beta t+\alpha +\beta st \\
		+\alpha -\beta+\alpha t+\beta t-\alpha -\beta st \\
		+\alpha -\beta -\alpha t-\beta t+\alpha +\beta st
	\end{bmatrix}
	\in S,\label{eqn:fstar}
\end{equation}
and one sees that e.g. the double point $[1,1,1,-1,-1,-1]$, which is on the line given by $(s,t)=(0,0)$, is not on the line given by $(s,t)=(0,1)$. Thus, these two lines are again distinct and so $\tilde{f_*}$ does not contract this particular exceptional divisor to a point.

Making the obvious changes, one shows in this way that none of the ten curves is contracted to a point. Hence, $D$ is isomorphic to $B_4$, a blow-up of the plane in 4 (rational) points. Moreover, the isomorphism $\tilde{f_*}$ gives an explicit parametrization of all of the lines on $S$, which can be used to find all chiral solutions to the anomaly cancellation conditions.

\subsubsection{The distribution of rational points}

del Pezzo surfaces of degree five are in many ways even simpler than cubic surfaces (which are del Pezzo surfaces of degree three), because they are obtained by blowing up fewer points in $\PP^2$.\footnote{A cubic surface over an algebraically-closed field, being a del Pezzo surface of degree three, is the blow up of $\PP^2$ at six points.} As such, it should come as no surprise that we can carry out a detailed study of the distribution of rational points, much as we did in \cite{Gripaios_2024_irreducible,Gripaios_2025_products}. In particular, it follows straightforwardly from our birational map that the rational points are dense in the Zariski topology and have the much stronger property of being dense in the real points in the usual Euclidean topology.\footnote{This follows because $\Q\PP^2$ is dense in $\mathbb{R}\PP^2$ in this topology and because our birational map fails to be defined only on the 10 exceptional curves.} In fact, in this case one can go somewhat further and give concrete results on the distribution of rational points of bounded height (cf. \cite{Gripaios_2024_irreducible,Gripaios_2025_products} where we discussed the case of cubic surfaces). Indeed we have that if we delete the 10 exceptional curves\footnote{These grow as $B^2$.} then the asymptotic growth of rational points of height $B$ is given by \cite{delaBreteche_2002,Browning_2022} 
\begin{gather}
	\frac{\pi^2}{72} \prod_p \left( 1-\frac{1}{p}\right)^5 \left( 1+\frac{5}{p} +\frac{1}{p^2}\right) B (\log B)^4
\end{gather}
thus verifying Manin’s conjecture \cite{Franke_1989} (and Peyre’s refinement thereof \cite{Peyre_1995}) in this case. This quantifies the qualitative observation that, even though there are infinitely-many integer solutions, there are few solutions where those charges are small.

We remark that, unlike in the case of solutions to the single $\mathfrak{u}_1$ case, where typically the chiral solutions dominate over the vectorlike ones, as measured by the dimension of a set in which they are dense, in this case, we get solutions which are equally prevalent, in that they are both dense in varieties of dimension two.
\subsubsection{Lines through points \label{sec:lp}} 
We close our discussion of this example, by using the results on lines in the Segre cubic primal $\mathfrak{S}$ to discuss the lines through a given fixed point $x \in \mathfrak{S}$. As well as allowing us to make contact with the approach of \cite{Costa_2020} and allowing us to give a transparent geometric picture of their results, it is of interest for physics {\em per se}. Indeed, one can easily imagine a phenomenologist finding themself in the situation of knowing the charges of a set of fermions under some $\mathfrak{u}_1$, and one wishing to find the possible charges (subject to the anomaly cancellation conditions) with respect to a second (perhaps putative) $\mathfrak{u}_1$.\footnote{A much studied example in physics beyond the Standard Model is the case of a possible additional $\mathfrak{u}_1^\prime$, where the fermion charges must be consistent with the hypercharges under the known Standard Model $\mathfrak{u}_1$; this case was solved in \cite{Allanach_2020_extragaugeboson}.}
In our approach, solving this problem precisely corresponds to finding all points on all lines through a given point $x \in \mathfrak{S}$. Clearly, once we know the lines, finding all points on them is a triviality, so we concentrate our efforts on finding the lines.

We have already shown in \cref{sec:failure_lines_through_point} that in general there need not be any lines through a given point, because in general we need to solve an arbitrary quadratic and an arbitrary cubic over the rationals. But as the algebraic approach of \cite{Costa_2020} showed, there are always six solutions in the case of pure $\mathfrak{u}_1\oplus \mathfrak{u}_1$ with six fermions.\footnote{This observation probably goes back to Segre.} As we will now see, these results (and more besides) can be easily recovered in our geometric approach.

To begin with, we observe that there cannot be exactly six lines through every point in $\mathfrak{S}$, because there are points on planes, and these obviously have infinitely many lines through them. So we must first identify the different types of point that can arise.

It turns out that the qualitative behaviour of the lines through $x$ is completely determined by whether or not $x$ is on a plane, and if so, whether or not it is a singular point.

Let us discuss the lines through $x \in \mathfrak{S}$ in each case. Whenever $x$ is on some plane,\footnote{We caution the reader that such an $x$ may lie on more that one plane. Indeed there are 45 lines in the variety corresponding to intersections of planes, and 15 regular points corresponding to intersections of a pair of such lines, i.e. intersections of triples of planes.} we clearly have all the lines through $x$ corresponding to lines in that plane through $x$; since these are non-chiral and easily found, we shall not discuss them further, restricting our discussion to the chiral lines.

Suppose first that $x$ is on 6 planes, i.e. is a singular point. As shown in Lemma 2.4 of \cite{Ciliberto_2024_modular}, there are no further lines through $x$ beyond those in the six planes.

Now suppose that $x$ is a regular point on some plane. There are now exactly 2 more lines beyond those in the planes (\emph{c.f.} Lemma 2.2 of \cite{Ciliberto_2024_modular}). To describe them explicitly, supposing for example that one of the planes is the image under $f$ of the plane $[123] \subset \PP^3$. Since $x$ cannot be a singular point, its preimage in  $\PP^3$ cannot be on any of the lines $\ell_{12},\ell_{13},\ell_{23},\ell_{45}$, since these are the preimages of the four singular points on the plane $f([123])$. There are then exactly two chiral lines through it, defined as the images under $f$ of the lines joining $g(x)$ to $p_4$ and $p_5$. (The other lines in the six components $D_\alpha$ get sent to lines in planes, so are not chiral.) 

Finally, suppose $x$ is generic, in that it lies on no plane.
We now show that there are exactly six lines through $x$, with one line in each of the six dP components that we have identified. To do so, we observe that not only is the map $g$ defined on such an $x$, but also that the image $g(x)$ is not on any of the 10 planes in $\PP^3$ going through any 3 of the special points $p_\alpha$ (since $g$ is invertible at $x$ and its inverse $f$ sends these planes in $\PP^3$ to planes in $\mathfrak{S}$). Five of the six lines in  $\mathfrak{S}$ through $x$ are then constructed by taking the five lines through $g(x)$ and each of the $p_\alpha$ and applying Richmond's map $f$. We have already seen that the image of such a line is a line, and it is a line through $x$ since $f$ inverts $g$ at $x$. The sixth line is the image under $f$ of the unique twisted cubic through the six points\footnote{Recall that we have previously shown that there is a 2-dimensional space of twisted cubics through five points in general position. By specifying that the curve passes through another point, we remove two degrees of freedom giving a unique curve.}  $g(x)$ and $p_\alpha$ (which are in general position), which we already know yields a line, and which again must go through $x$, since $f$ inverts $g$ there. From our characterization of all the lines in the variety, it is clear that there can be no other lines through such an $x$.

For the purposes of finding actual values of the charges one needs formul\ae , so let us indicate how these can be found. Given $x = [x_0:x_1:x_2:x_3:x_4:x_5] \in\mathfrak{S}$, we get that $g(x) = [a:b:c:d] := [
+(x_0+x_4)(x_1+x_3)(x_2+x_5) :
-(x_0+x_4)(x_1+x_3)(x_3+x_4) :
-(x_1+x_2)(x_1+x_3)(x_2+x_5) :
-(x_0+x_4)(x_0+x_5)(x_2+x_5)]
\in\PP^3$.
We claim that the five lines through $x$ that are in the five del Pezzo components $D_1,\dots,D_5$ are defined by the fact that they pass through the following other points on $\mathfrak{S}$:
\begin{align}
	D_1:
	\begin{bmatrix}
		-b-c+d \\
		+b-c+d \\
		-b+c-d \\
		+b-c-d \\ 
		-b+c+d \\
		+b+c-d
	\end{bmatrix},  
	D_2:
	\begin{bmatrix}
		-a+c-d \\
		+a+c-d \\
		-a+c+d \\
		+a-c-d \\
		-a-c+d \\
		+a-c+d
	\end{bmatrix}, 
	D_3:
	\begin{bmatrix}
		-a+b+d \\
		-a+b-d \\
		+a+b-d \\
		-a-b+d \\
		+a-b-d \\
		-a-b+d
	\end{bmatrix}, \nonumber \\
	D_4:
	\begin{bmatrix}
		+a-b+c \\
		+a-b-c \\
		-a+b-c \\
		-a+b+c \\
		+a+b-c \\
		-a-b+c
	\end{bmatrix}, 
	D_5:
	\begin{bmatrix}
		-a-b+c+d \\
		+a+b-c-d \\
		-a+b+c-d \\
		-a+b-c+d \\
		+a-b-c+d \\
		+a-b+c-d
	\end{bmatrix}.
\end{align}
To prove the claim for $D_1$, for example, recall that the preimage in $\PP^3$ under $f$ of the line in $D_1$ through $x$ is a line through $p_1$ and $g(x)$ and so can be written as $[a':b':c':d']=[as+t : bs : cs : ds]$ for $[s:t]\in\PP^1$. Under Richmond's map $f$, this line gets sent to the line $xs+
\begin{bmatrix}
	-b-c+d \\
	+b-c+d \\
	-b+c-d \\
	+b-c-d \\
	-b+c+d \\
	+b+c-d	
\end{bmatrix}t$, (where we have ignored an overall multiplicative factor of $s$), which evidently goes through the claimed point. 

A formula for the sixth line through $D_0$ can be found as follows.
We have an expression \cref{eq:twist} for (an affine open subset of) the twisted cubics through the five points $p_\alpha \in \PP^3$ and we wish to find the unique twisted cubic through a sixth given point $[a:b:c:d]$. Solving, we get
\begin{gather}
	u=\frac{d(a-b)}{b(a-d)}, v=\frac{d(a-c)}{c(a-d)}, t=\left(1-\frac{a}{d}\right)s
\end{gather}
which fixes both the cubic and the point of intersection. These values of $u$ and $v$ can be plugged into the expression in \cref{eqn:D0} to obtain two points lying on this sixth line, which also passes through $x$ when $t$ and $s$ are related as above.

There is one more issue relevant for physics that we should discuss. As we discussed at the end of the introduction, \cite{Costa_2020} observed in this example that there exist solutions which are non-chiral with respect to each $\mathfrak{u}_1$ individually, but nevertheless chiral overall (because the pairing of Weyl fermions into Dirac fermions in each case by a permutation). This corresponds more generally to our observation in \cref{sec:alggeo} that chiral lines may, despite their name, occasionally contain pairs of points that, either individually or collectively, correspond to 
non-chiral solutions.

In this particular example, these solutions must correspond to intersections of chiral lines with the 15 planes in $\mathfrak{S}$. So to understand these solutions, we need to study such intersections.

We claim that every chiral line passes through exactly five planes. Indeed, every chiral line is on some del Pezzo component, and since all of these components are isomorphic by permutations of $x_i$, we can without loss of generality consider a line $L$ in $D_1$. Such an $L$ must also go through a generic $x$ if it is to be chiral, so we can characterize it uniquely as the image under $f$ of the line $L^\prime$ joining $p_1$ and $g(x)$ in $\PP^3$.
This line $L^\prime$ intersects each of the four planes $[\alpha \beta \gamma]$ with $2\leq \alpha<\alpha<\alpha\leq 5$ in $\PP^3$ that are disjoint from $p_1$ in a point and the images of each these four intersection points are contained in the corresponding planes $f([\alpha \beta \gamma]) \in \mathfrak{S}$. Moreover, since $g(L)$ passes through $p_1$, $L$ will also intersect a fifth plane, namely the image $\tilde{f}(E_1)$ of the exceptional divisor at the point $p_1$ in the blow up $B_5$.

For completeness, we specify which 5 planes are intersected by each of the six (chiral) lines through a generic $x \mathfrak{S}$. For the lines through $D_2,\dots,D_5$, this is done exactly as we have just done for $D_1$. For $D_0$ we observe that the five planes are given by the images under $\tilde{f}$ of all five exceptional divisors $E_\alpha$ (since the twisted cubic through $x$ goes through all five $p_\alpha$).

Using our knowledge of the preimages of the fifteen planes on $\mathfrak{S}$ under $f$ (\emph{c.f.} \cref{eqn:planes1,eqn:planes2}), we can summarise the results as follows: a line through a general point $x$ on $\mathfrak{S}$ will intersect
\begin{enumerate}
	\item $(05|12|34),(03|14|25),(04|15|23),(02|13|45),(01|24|35)$ if it lies on $D_0$,
	\item $(05|12|34),(04|13|25),(03|15|24),(02|14|35),(01|23|45)$ if it lies on $D_1$,
	\item $(03|14|25),(02|15|34),(04|12|35),(05|13|24),(01|23|45)$ if it lies on $D_2$,
	\item $(04|15|23),(03|12|45),(01|25|34),(05|13|24),(02|14|35)$ if it lies on $D_3$,
	\item $(02|13|45),(05|14|23),(01|25|34),(04|12|35),(03|15|24)$ if it lies on $D_4$,
	\item $(01|24|35),(05|14|23),(03|12|45),(02|15|34),(04|13|25)$ if it lies on $D_5$.
\end{enumerate}

It follows that a generic chiral line will yield $10$ solutions of this type to the $\mathfrak{u}_1 \oplus \mathfrak{u}_1$ anomaly cancellation conditions, corresponding the the choice of five unordered pairs of intersections of the line with the planes. But we stress that not all chiral lines will have exactly 10 solutions, because we know that some chiral lines intersect more than one plane at the same point, namely where two planes intersect in a line.\footnote{We have seen that there are no chiral lines through singular points, and indeed this is consistent with what we have also just seen, namely that such lines go through exactly five, ergo not six, planes.} Moreover, some of these solutions may have sufficiently many charges the same that they are not chiral overall.

\section{Discussion \label{sec:otherex}}
An obvious question is whether our methods can be expected to work for other examples with a
$\mathfrak{u}_1 \oplus \mathfrak{u}_1$ summand. These all correspond to studying lines in cubic hypersurfaces, which has become, starting with e.g. \cite{Altman_1977}, a well-studied area in algebraic geometry.

A first observation is that almost none of the examples we studied actually correspond to rational varieties per se. Indeed, for the cubic surfaces in \cref{sec:cubic_surface}, the variety of lines is zero-dimensional and has up to 27 points. It is therefore (unless it is a single point) a reducible variety, whose irreducible components are rational varieties. While for the Segre cubic primal in \cref{sec:u1u1}, we get a variety which is highly reducible, but each of whose irreducible components is a rational surface.

This irrationality of the Fano variety of lines turns out to be a general feature of low dimensional cubic hypersurfaces, at least if they are smooth. Indeed, for $d \leq 4$, the Fano variety of a lines of a smooth cubic $d$-fold hypersurface has non-negative Kodaira dimension, so is necessarily irrational. Moreover, it is necessarily smooth, so for $d>2$ it cannot be reduced into rational irreducible components.

We thus expect the cases corresponding to smooth cubic 3- or 4-folds to be much more challenging. For 3-folds, we get a surface of general type and so the Bombieri-Lang conjecture \cite{Vojta_1987} suggests that the rational points should not even be Zariski dense. So solutions to the anomaly cancellation equations are likely to be few and far between, and much harder to find. For smooth cubic 4-folds (at least over an algebraically-closed field) the Fano variety of lines is a hyperk\"ahler fourfold \cite{Beauville_1985} and here too much progress has been made, though of a rather different nature (see e.g. \cite{Huybrechts_2024,Galkin_2014}).

In particular, it follows that the next `obvious' example, of pure $U(1)^2$ gauge theory with seven (rather than six) fermions, corresponding to the smooth \emph{Clebsch-Segre cubic fourfold} defined by
\begin{equation}
	x_0^3+x_1^3+x_2^3+x_3^3+x_4^3+x_5^3+x_6^3 = x_0+x_1+x_2+x_3+x_4+x_5+x_6=0 \label{eqn:clebsch_segre_fourfold}
\end{equation}
for $[x_0:x_1:x_2:x_3:x_4:x_5:x_6]\in\PP^6$ has a Fano variety of lines that is a smooth (in particular irreducible) irrational 4-fold. The underlying cubic 4-fold (which generalizes the smooth Clebsch cubic surface and the singular Segre cubic primal threefold in an obvious way) contains 105 planes, given by $x_0+x_1=x_2+x_3=x_4+x_5=0$ and its orbit under the obvious action of $S_7$ on the hypersurface \cite{Degtyarev_2023}. The lines in these planes, which correspond to non-chiral solutions of the anomaly cancellation conditions, form subvarieties, each isomorphic to $\PP^3$, of codimension one of the Fano variety of lines. So from dimensional considerations alone, there should be many more chiral solutions than non-chiral ones, but if the Bombieri-Lang conjecture is true, there are likely to be fewer. Evidently we have at least families of rational chiral solutions of codimension 2, given by taking solutions to the case with 6 fermions and setting the remaining charge to zero, but to our knowledge no other chiral solutions have been found. 

For cubic hypersurfaces of higher dimension, it is possible for the Fano variety of lines to be rational, but much less is known in general.

{\em Acknowledgements.}
BG thanks Mark Gross and Dhruv Ranganathan for discussions. This work was partially supported by STFC consolidated grants ST/T000694/1 and ST/X000664/1 and a Trinity-Henry Barlow Scholarship.

\appendix

\section{Fano varieties of linear subspaces} \label{app:fano}

Here we turn to the task of describing elements of the Fano variety $F_k(X)$ of $k$-planes on a projective hypersurface $X\in\PP^n$ explicitly. For this, we follow the algorithm outlined in \cite{Eisenbud_2016}, beginning by finding a cover of the grassmannian $\mathbb{G}(k,n)$ by affine open sets. One such set $U$ is given by the set of all $k$-planes not intersecting a fixed $(n-k-1)$-plane $\Pi$ in $\PP^n$. In coordinates where $\Pi$ is given by the vanishing of the first $k+1$ coordinates, then $U$ can be identified with the affine space of $(k+1)\times(n-k)$ matrices, such that any $k$-plane $L\in U$ is the row space of a unique matrix of the form

\begin{equation}
	A=
	\begin{pmatrix}
		1 & 0 & \dots & 0 & a_{0,k+1} & \dots & a_{0,n} \\
		0 & 1 & \dots & 0 & a_{1,k+1} & \dots & a_{1,n} \\
		\vdots & \vdots & \ddots & \vdots & \vdots & \ddots & \vdots \\
		0 & 0 & \dots & 1 & a_{k,k+1} & \dots & a_{k,n}
	\end{pmatrix}. \label{appeqn:matrix_A}
\end{equation}
This allows us to give a parameterization $h:\PP^k\to L$ of $L$ as
\begin{equation}
	\PP^k\ni [s_0:\dots:s_k] \xmapsto{h} (s_0\,\dots\, s_k)A = \left[s_0:\dots:s_k:\sum_{i=0}^k a_{i,k+1}s_i: \dots: \sum_{i=0}^k a_{i,n} s_i\right],
\end{equation}
where the $a_{i,j}$ are coordinates on $U$, and the $s_i$ are homogeneous coordinates on $\PP^k$. 

If $X\subset\PP^n$ is the hypersurface $V(g)$ given by a homogeneous degree $d$ polynomial in the $n+1$ homogeneous coordinates $z_0,\dots,z_n$ on $\PP^n$, then by substituting the parameterization $h$ of the line $L$ into $g$, we get a homogeneous polynomial of degree $d$ in the $k+1$ variables $s_0,\dots,s_k$ on $\PP^k$, whose coefficients are polynomials in the coordinates $\{a_{i,j}\}$ on $U\subset\mathbb{G}(k,n)$. The simultaneous vanishing of all of these coefficients serve as defining equations for $F_k(X)$ on this affine open set.

For the examples of lines in cubic surfaces considered in \cref{sec:cubic_surface}, we find it more convenient to use a different approach, based on the \emph{Pl\"{u}cker embedding} of a grassmannian in projective space. Recall that $\mathbb{G}(k,n)$ as we have defined it is equivalently the space of $k+1$-dimensional vector subspaces of a vector space $V$ of dimension $n+1$. Such a subspace is defined by $k+1$ linearly-independent vectors $w_1,\dots,w_{k+1}$ whose exterior product $w_1 \wedge \dots \wedge w_{k+1} \in \bigwedge^{k+1}V$ defines a point in $\PP(\bigwedge^{k+1}V)$ that is independent of the particular choice of linearly-independent vectors. Hence we have a well-defined embedding $G(k,n) \hookrightarrow \PP(\bigwedge^{k+1}V)$.

To get co-ordinates on this embedding, choose a basis $\{e_1, \dots, e_{n+1}\}$ for $V$. Then $\{e_{i_1} \wedge \dots \wedge e_{i_k+1}| 1 \leq i_1 \leq \dots \leq i_k+1 \leq n+1\}$ is a basis for $\bigwedge^{k+1}V$ and so we may expand $w_1 \wedge \dots \wedge w_{k+1} = \sum_I p_I e_I$, where $I = (i_1, \dots, i_{n+1})$.
The $p_I$ are called \emph{Pl\"{u}cker co-ordinates}. To relate them to our previous co-ordinates on the grassmannian, we observe that they are given by the $(k+1) \times (k+1)$-minors of the matrix $A$ in \cref{appeqn:matrix_A}.

Specializing now to lines on cubic surfaces in $\PP^3$, we have that $F_1(X)$ is zero-dimensional and so consists of a finite set of points. So there is an easy strategy to fully determine it: find all of its points. An algorithm for doing so is presented in \cite{Boissiere_2007}, which we reproduce here. 

As already described, a \emph{general} line $L\in\mathbb{G}(1,3)$ can be parameterized by a generic rank-two matrix of the form
\begin{equation}
	A=\begin{pmatrix}
		a & b & c & d \\
		e & f & g & h
	\end{pmatrix}.
\end{equation}
and the Pl\"ucker coordinates are the $2\times 2$-minors of this matrix:
\begin{align}
	p_{12}:=af-be,&&p_{13}:=ag-ce,&&p_{14}:=ah-de,\nonumber \\
	p_{23}:=bg-cf,&&p_{24}:=bh-df,&&p_{34}:=ch-dg.
\end{align}

In these coordinates, the embedding $\mathbb{G}(1,3)\hookrightarrow \PP^5$ corresponds to the hypersurface $p_{12}p_{34}-p_{13}p_{24}+p_{14}p_{23}=0$. 
Now consider the stratification of the grassmannian defined by the Pl\"ucker strata
\begin{enumerate}
	\item $p_{12}=1: A=\begin{pmatrix}
		1 & 0 & c & d \\ 0 & 1 & g & h
	\end{pmatrix}$.
	\item $p_{12}=0, p_{13}=1: A = \begin{pmatrix}
		1 & b & 0 & d \\ 0 & 0 & 1 & h
	\end{pmatrix}$.
	\item $p_{12}=p_{13}=0, p_{14}=1: A = \begin{pmatrix}
		1 & b & c & 0 \\ 0 & 0 & 0 & 1
	\end{pmatrix}$.
	\item $p_{12}=p_{13}=p_{14}=0, p_{23}=1: A = \begin{pmatrix}
		0 & 1 & 0 & d \\ 0 & 0 & 1 & h
	\end{pmatrix}$.
	\item $p_{12}=p_{13}=p_{14}=p_{23}=0, p_{24}=1: A = \begin{pmatrix}
		0 & 1 & c & 0 \\ 0 & 0 & 0 & 1
	\end{pmatrix}$.
	\item $p_{12}=p_{13}=p_{14}=p_{23}=p_{24}=0, p_{34}=1: A = \begin{pmatrix}
		0 & 0 & 1 & 0 \\ 0 & 0 & 0 & 1
	\end{pmatrix}$.
\end{enumerate}

The first stratum is evidently the affine open subset of $\mathbb{G}(1,3)$ that contains all lines not intersecting the line $x_0=x_1=0$, while the subsequent strata cover its complement. Thus, we can find all lines on a cubic surface by pulling back the equation of a line in each Pl\"ucker stratum to the cubic surface and requiring that the resulting polynomial vanish identically. This can be easily done using computer algebra packages that can handle rational numbers, such as \texttt{Mathematica}.

\section{Example of finding lines in a cubic surface \label{app:clebsch_lines}}

We illustrate the method described in \cref{app:fano} by finding the equations of all of the rational lines on the Clebsch diagonal cubic surface
\begin{equation}
	x_0(x_1+x_2+x_3)(x_0+x_1+x_2+x_3)+(x_1+x_2)(x_1+x_3)(x_2+x_3) = 0.
\end{equation}

A line in the first Pl\"ucker stratum is parameterized by
\begin{equation}
	[x_0:x_1:x_2:x_3]=[s_0:s_1:cs_0+gs_1:ds_0+hs_1],
\end{equation}
and so it will lie on the Clebsch diagonal cubic surface if the equation
\begin{multline}
	s_0[(c+d)s_0+(1+g+h)s_1][(1+c+d)s_0+(1+g+h)s_1]\\
	+[cs_0+(1+g)s_1][ds_0+(1+h)s_1][(c+d)s_0+(g+h)s_1]=0
\end{multline}
is satisfied for all values of $[s_0:s_1]\in\PP^1$. This gives rise to four simultaneous equations that respectively correspond to the vanishing of the coefficients of $s_0^3$, $s_0^2s_1$, $s_0s_1^2$ and $s_1^3$ on the left-hand side:
\begin{align}
	(1+c)(1+d)(c+d)&=0, \\
	c^2g+d^2h-(1+c+d)^2(1+g+h)&=0, \\
	cg^2+dh^2-(1+c+d)(1+g+h)^2&=0, \\
	(1+g)(1+h)(g+h)&=0.
\end{align}

These equations can be solved over $\Q$, where we find six rational lines with $(c=h=0, d=g=-1)$, $(c=h=-1, d=g=0)$, $(c=-1, d=g=h=0)$, $(d=-1, c=g=h=0)$, $(g=-1, c=d=h=0)$ and $(h=-1, c=d=g=0)$. These correspond to the lines $x_0+x_3=x_1+x_2=x_4=0$, $x_0+x_2=x_1+x_3=x_4=0$, $x_0+x_2=x_1+x_4=x_3=0$, $x_0+x_3=x_1+x_4=x_2=0$, $x_0+x_4=x_1+x_2=x_3=0$ and $x_0+x_4=x_1+x_3=x_2=0$.

In the second Pl\"ucker stratum, a line can be parameterized by
\begin{equation}
	[x_0:x_1:x_2:x_3]=[s_0:bs_0:s_1:ds_0+hs_1].
\end{equation}
Now, the equations to be solved are
\begin{align}
	(1+b)(1+d)(b+d)&=0, \\
	d^2h-(1+b+d)^2(1+h)&=0, \\
	dh^2-(1+b+d)(1+h)^2&=0, \\
	h(1+h)&=0.
\end{align}
whose solutions (all rational) are: $(b=h=-1, d=0)$, $(b=-1, d=h=0)$, $(d=-1, b=h=0)$ and $(h=-1, b=d=0)$. These correspond to the rational lines $x_0+x_1=x_2+x_3=x_4=0$, $x_0+x_1=x_2+x_4=x_3=0$, $x_0+x_3=x_2+x_4=x_1=0$ and $x_0+x_4=x_2+x_3=x_1=0$.

In the third stratum, a line is parameterized by
\begin{equation}
	[x_0:x_1:x_2:x_3]=[s_0:bs_0:cs_0:s_1],
\end{equation}
and it lies on the Clebsch diagonal cubic surface if
\begin{equation}
	(1+b)(1+c)(b+c)=(1+b+c)=0.
\end{equation}
These equations have two solutions $(b=-1,c=0)$ and $(b=0,c=-1)$, which correspond to the two rational lines $x_0+x_1=x_2=x_3+x_4=0$ and $x_0+x_2=x_1=x_3+x_4=0$. 

In the fourth stratum, a line is parameterized by
\begin{equation}
	[x_0:x_1:x_2:x_3]=[0:s_0:s_1:ds_0+hs_1],
\end{equation}
and it lies on the surface if
\begin{equation}
	d(1+d)=h+2dh+(1+d)^2=d+2dh+(1+h)^2+h(1+h)=0.
\end{equation}
The solutions are $(d=-1,h=0)$ and $(d=0,h=-1)$, which correspond to the two rational lines $x_0=x_1+x_3=x_2+x_4=0$ and $x_0=x_1+x_4=x_2+x_3=0$.

In the fifth stratum, a line is parameterized by
\begin{equation}
	[x_0:x_1:x_2:x_3]=[0:s_0:cs_0:s_1],
\end{equation}
and it lies on the surface if
\begin{equation}
	c(1+c)=1+c=0.
\end{equation}
The only solution is obviously $c=-1$, giving the fifteenth rational line $x_0=x_1+x_2=x_3+x_4=0$. 

Finally, it is straightforward to check that no line in the sixth Pl\"ucker stratum, which has the form
\begin{equation}
	[x_0:x_1:x_2:x_3]=[0:0:s_0:s_1]
\end{equation}
can lie entirely on the Clebsch diagonal cubic surface.

\section{A cover of the grassmannian of lines on $\PP^4$} \label{app:cover}
Here we show that the grassmannian $\mathbb{G}(1,4)$ of projective lines in the $\PP^4 \subset \PP^5$ defined by $x_0+x_1+x_2+x_3+x_4+x_5 = 0$  is covered by the fifteen affine open sets $U_{ij,kl,mn}$ consisting of lines that do not intersect the plane $(ij|kl|mn)$ (defined by $x_i+x_j=x_k+x_l=x_m+x_n=0$ for $\{i,j,k,l,m,n\}=\{0,1,2,3,4,5\}$).\footnote{Note that Example 4.8 and Proposition 4.10 of \cite{Roth_2004} show that at least 7 affine charts are required.}

To do so, we must show that no line intersects all 15 planes. Any line may be written as $(x_0(s,t),x_1(s,t), \dots, x_6(s,t))$, where $[s:t] \in \PP^1$ is a parameter along the line and we may regard the $x_i(s,t)$ as linear forms on the 2-dimensional vector space defined by $\{(s,t)\}$. In order for this line to intersect the plane $(ij|kl|mn)$, there must exist an $(s,t)$ such that $x_i(s,t)+x_j(s,t)=x_k(s,t)+x_l(s,t)=x_m(s,t)+x_n(s,t)=0$. In other words, the 3 linear forms $x_i(s,t)+x_j(s,t), x_k(s,t)+x_l(s,t), x_m(s,t)+x_n(s,t)$ must be proportional.

Now, at least one of the forms of the form $x_i(s,t)+x_j(s,t)$ (considering all values of $i,j \in \{0,1,2,3,4,5\}$ with $i \neq j$) must be a non-zero form. (For if not, all of the forms $x_i (s,t)$ would have to be the zero form, contradicting the assumption that they define a line.) Let us suppose that the form $x_0(s,t) + x_1(s,t)$ is non-zero and given by $x(s,t) \neq 0$. Evidently, if instead one of the other forms $x_i(s,t)+x_j(s,t)$ is non-zero, we can re-run the argument that follows simply by approriately permuting the indices.

Now consider the 3 planes $(01|23|45), (01|24|35), (01|25|34)$. Since our line must intersect all three planes, we conclude that the forms $x_2(s,t)+x_3(s,t)$, $x_4(s,t)+x_5(s,t)$, $x_2(s,t)+x_4(s,t)$, $x_3(s,t)+x_5(s,t)$, $x_2(s,t)+x_5(s,t)$, and  $x_3(s,t)+x_4(s,t)$ are all proportional to $x(s,t)$. By taking linear combinations, it follows that $x_2(s,t)$, $x_3(s,t)$, $x_4(s,t)$, and $x_5$ are all proportional to $x(s,t)$.

Next consider a 4th plane, viz. $(0j|1l|mn)$ with $j,l,m,n$ taking distinct values in $\{2,3,4,5\}$. For our line to hit it, we must have that $x_0(s,t)+x_j(s,t)$, $x_1(s,t)+x_l(s,t)$, and $x_m(s,t)+x_n(s,t)$ are proportional. Now comes the crux of the argument. As long as $x_m(s,t)+x_n(s,t)$ is not the zero form, it follows from the considerations in the last paragraph that $x_0(s,t)+x_j(s,t)$ and $x_1(s,t)+x_l(s,t)$ are also proportional to $x(s,t)$, whence $x_0(s,t) = (x_0(s,t) +x_j(s,t)) - x_j(s,t)$ and similarly $x_1(s,t)$ are proportional to $x(s,t)$. So all six forms $x_i(s,t)$ are proportional, which means they describe a point in projective space, contradicting the starting assumption that they define a line.

This argument is watertight unless $x_m(s,t)+x_n(s,t)$ is the zero form. If it is, consider some other choice of distinct $m,n \in \{2,3,4,5\}$ and run the same argument. One of these pairs must work unless $x_m(s,t)+x_n(s,t)$ is always the zero form. But if it is then $x_m$ is the zero form for every $m \in \{2,3,4,5\}$ and the relation $\sum_i x_i = 0$ implies that $x_0(s,t)+x_1(s,t)$ is also the zero form, which is again a contradiction. So no line intersects these six planes, and no line intersects all 15 planes.

\section{Richmond's map is a birational equivalence} \label{app:inverse} 
We show that, when they are well-defined, the following maps are left- and right-inverses of each other on some affine open set:
\begin{equation}
	f:\PP^3\ni[a:b:c:d]\mapsto
	\begin{bmatrix}
		x_0 \\ x_1 \\ x_2 \\ x_3 \\ x_4 \\ x_5
	\end{bmatrix}
	:=
	\begin{bmatrix}
		-ac+bc+ad-bd-ab+cd \\
		-ab+bc+ad-bd+ab-cd \\
		+ac+bc-ad+bd-ab-cd \\
		-ac-bc-ad+bd+ab+cd \\
		+ac-bc+ad+bd-ab-cd \\
		+ac-bc-ad-bd+ab+cd
	\end{bmatrix}
	\in\mathfrak{S},
\end{equation}
\begin{equation}
	g:\mathfrak{S}\ni
	\begin{bmatrix}
		x_0 \\ x_1 \\ x_2 \\ x_3 \\ x_4 \\ x_5
	\end{bmatrix}
	\mapsto
	\begin{bmatrix}
		a \\ b \\ c \\ d
	\end{bmatrix}
	:=
	\begin{bmatrix}
		(x_0+x_4)(x_1+x_3)(x_2+x_5) \\
		-(x_0+x_4)(x_1+x_3)(x_3+x_4) \\
		-(x_1+x_2)(x_1+x_3)(x_2+x_5) \\
		-(x_0+x_4)(x_0+x_5)(x_2+x_5)
	\end{bmatrix}
	\in \PP^3
\end{equation}
where
\begin{equation}
	x_5=-(x_0+x_1+x_2+x_3+x_4)
\end{equation}
and $\mathfrak{S}$ is the Segre cubic primal defined by
\begin{equation}
	x_0^3+x_1^3+x_2^3+x_3^3+x_4^3-(x_0+x_1+x_2+x_3+x_4)^3=0.
\end{equation}
On the one hand, we have:
\begin{align}
	g\circ f:[a:b:c:d]&\xmapsto{f}
	\begin{bmatrix}
		x_0 \\ x_1 \\ x_2 \\ x_3 \\ x_4 \\ x_5 
	\end{bmatrix}
	&= 
	\begin{bmatrix}
		-ac+bc+ad-bd-ab+cd \\
		-ac+bc+ad-bd+ab-cd \\
		ac+bc-ad+bd-ab-cd \\
		-ac-bc-ad+bd+ab+cd \\
		ac-bc+ad+bd-ab-cd \\
		ac-bc-ad-bd+ab+cd
	\end{bmatrix} \nonumber \\
	&\xmapsto{g}\begin{bmatrix}
		a' \\ b' \\ c' \\d'
	\end{bmatrix}&=\begin{bmatrix}
		-8a^3(b-c)(b-d)(c-d) \\ -8a^2b(b-c)(b-d)(c-d) \\ -8a^2c(b-c)(b-d)(c-d) \\ -8a^2d(b-c)(b-d)(c-d)
	\end{bmatrix} \nonumber \\
	& &= [a:b:c:d]
\end{align}
when $a^2(b-c)(b-d)(c-d) \neq 0$, so $g\circ f = 1$ on an affine open set.

On the other hand, to evaluate $f\circ g$, we will need that the equation of the Segre cubic primal implies
\begin{equation}
	\sum_{\substack{i\neq j \\ i,j\in\{0,1,2,3,4\}}}x_i^2x_j + \sum_{\substack{i<j<k \\ i,j,k\in\{0,1,2,3,4\}}}2x_ix_jx_k=0.\label{eqn_app:segre_special}
\end{equation}
Then (suppressing the coordinate $x_5$),
\begin{align}
	f\circ g:
	\begin{bmatrix}
		x_0 \\ x_1 \\ x_2 \\ x_3 \\ x_4
	\end{bmatrix}
	\xmapsto{g}[a:b:c:d]
	&=\begin{bmatrix}
		(x_0+x_4)(x_1+x_3)(x_0+x_1+x_3+x_4) \\
		(x_0+x_4)(x_1+x_3)(x_3+x_4) \\
		-(x_1+x_2)(x_1+x_3)(x_0+x_1+x_3+x_4) \\
		(x_0+x_1+x_3+x_4)(x_0+x_4)(x_1+x_2+x_3+x_4)
	\end{bmatrix} \nonumber \\
	&\xmapsto{f} (x_0+x_4)(x_1+x_3)(x_0+x_1+x_3+x_4)[x'_0:x'_1:x_2':x_3':x'_4] \nonumber \\
	&=[x_0':x_1':x_2':x_3':x_4'],
\end{align}
where we used \cref{eqn_app:segre_special} to evaluate that, for $i\in\{0,1,2,3,4\}$,
\begin{equation}
	x_i'=2x_i\left(x_1^2 + x_4^2 + \sum_{\substack{i\neq j \\ i,j\in\{0,1,2,3,4\}}}x_i x_j\right)\propto x_i,
\end{equation}
such that indeed $f\circ g= 1$ on an affine open set.

\section{Matching del Pezzo components} \label{app:dP}

Here we will use the map $g$ to identify the (affine open subsets of the) non-linear irreducible components of the Fano variety of lines on the Segre cubic primal, as described in \cref{sec:segre_cover}, with the del Pezzo surfaces found in \cref{sec:induced_maps}.
\subsection{The affine patch $U_{01,23,45}$}
Consider a line on the affine patch $U_{01,23}$ of the Fano variety of lines $F_1(\mathfrak{S})$ on the Segre cubic primal $\mathfrak{S}$ parameterized by 
\begin{equation}
	U_{01,23}\ni x=
	\begin{bmatrix}
		x_0 \\ x_1 \\ x_2 \\ x_3 \\ x_4 \\ x_5
	\end{bmatrix}
	=
	\begin{bmatrix}
		u(1+a_2)+v b_2 \\
		u(1-a_2)-v b_2 \\
		u a_3+v(1+b_3) \\
		-u a_3 + v(1-b_3) \\
		u a_4 + v b_4 \\
		-u(2+a_4)-v(2+b_4)
	\end{bmatrix}
\end{equation}
for $[u:v]\in\PP^1$. Let us write $a_{ij}^\pm$ for $a_i\pm a_j$, and similarly for $b_{ij}^\pm$. On this patch, four non-linear components of the Fano variety of lines on the Segre cubic primal, $F_1(\mathfrak{S})$, are visible:
\begin{enumerate}
	\item The component with $a_{24}^+=b_{34}^+=-1, a_{23}^+=b_{23}^+, (a_{23}^+)^2=4a_2b_3+1$.
	
	If $b_3\neq0$, then we can choose the free variables to be $b_2$ and $b_3$, for which $a_2=\frac{(b_{23}^+)^2-1}{4b_3}$ and $a_3=b_{23}^+-a_2$. Then, we have:
	\begin{equation}
		\begin{bmatrix}
			u(1+a_2)+v b_2 \\
			u(1-a_2)-v b_2 \\
			u a_3+v(1+b_3) \\
			-u a_3+v(1-b_3) \\
			-u(a_2+1)-v(b_3+1) \\
			u(a_2-1)+v(b_3-1)
		\end{bmatrix}
		\xmapsto{g}
		\begin{bmatrix}
			v(u+v)[(b_{23}^+-1)u+2b_3v] \\
			v(u+v)[(b_{23}^+u+2b_3v] \\
			\dfrac{1}{2b_3}(u+v)[(b_{23}^+-1)u+2b_3v][(b_{23}^++1)u+2b_3v] \\
			\dfrac{1}{2b_3}v[(b_{23}^+-1)u+2b_3v][(b_{23}^++1)u+2b_3v]
		\end{bmatrix},
	\end{equation}
	if $(b_{23}^--1)(b_{23}^+-1)\neq0$ such that we can divide by the latter. This is a two-parameter family of twisted cubics passing through the following five points on $\PP^3$: $p_1=[1:0:0:0]$ (when $[u:v]=[-2b_3:(b_{23}^++1)]$), $p_2=[0:1:0:0]$ (when $[u:v]=[-2b_3:(b_{23}^+-1)]$), $p_3=[0:0:1:0]$ (when $[u:v]=[1:0]$), $p_4=[0:0:0:1]$ (when $[u:v]=[1:-1]$) and $p_5=[1:1:1:1]$ (when $[u:v]=[0:1]$). This is a part of the del Pezzo component $D_0$ that is visible in the affine patch $U_{01,23,45}$.
	
	\item The component with $a_{24}^+=-b_{34}^-=-1, a_{23}^-=b_{23}^-, (a_{23}^-)^2=-4a_2b_3+1$.
	
	If $b_3\neq0$, then we can choose the free variables to be $b_2$ and $b_3$, for which $a_2=\frac{1-(b_{23}^-)^2}{4b_3}$ and $a_3=a_2-b_{23}^-$. Then, we have
	\begin{equation}
		\begin{bmatrix}
			u(1+a_2)+v b_2 \\
			u(1-a_2)-v b_2 \\
			u a_3+v(1+b_3) \\
			-u a_3+v(1-b_3) \\
			-u(1+a_2)+v(b_3-1) \\
			u(a_2-1)-v(1+b_3)
		\end{bmatrix}
		\xmapsto{g}
		u
		\begin{bmatrix}
			0 \\ 0 \\ 1 \\ 0
		\end{bmatrix}
		+v
		\begin{bmatrix}
			\dfrac{b_{23}^+-1}{b_{23}^--1} \\
			\dfrac{b_{23}^++1}{b_{23}^-+1} \\
			1 \\
			1
		\end{bmatrix},
	\end{equation}
	if the factor $\frac{(b_{23}^-+1)(b_{23}^--1)(b_{23}^+-1)^2[(b_{23}^-+1)u-2b_3v]u}{4b_3^2}$ is non-vanishing such that we can divide by it. Thus, a general line is sent by $g$ to a line on $\PP^3$ that passes through $p_3=[0:0:1:0]$ (when $[u:v]=[1:0]$) and $\left[\frac{b_{23}^+-1}{b_{23}^--1}:\frac{b_{23}^++1}{b_{23}^-+1}:0:1\right]$ (when $[u:v]=[1:-1]$). (N.B. We cannot set $u=0$ in the above expression because then the factor that we are dividing out vanishes.) This is a part of the del Pezzo component $D_3$ that is visible in the affine patch $U_{01,23,45}$.
	
	\item The component with $a_{24}^-=b_{34}^-=1, a_{23}^+=b_{23}^+, (a_{23}^+)^2=4a_2b_3+1$.
	
	If $b_3\neq0$, then we can choose the free variables to be $b_2$ and $b_3$, for which $a_2=\frac{(b_{23}^+)^2-1}{4b_3}$ and $a_3=b_{23}^+-a_2$. Then, we have:
	\begin{equation}
		\begin{bmatrix}
			u(1+a_2)+v b_2 \\
			u(1-a_2)-v b_2 \\
			u a_3+v(1+b_3) \\
			-u a_3+v(1-b_3) \\
			u(a_2-1)+v(b_3-1) \\
			-u(a_2+1)-v(b_3+1)
		\end{bmatrix}
		\xmapsto{g}
		v
		\begin{bmatrix}
			0 \\ 0 \\ 0 \\ 1
		\end{bmatrix}
		+(u+v)
		\begin{bmatrix}
			\dfrac{b_{23}^+-1}{b_{23}^--1} \\
			\dfrac{b_{23}^++1}{b_{23}^-+1} \\
			1 \\ 
			0
		\end{bmatrix},
	\end{equation}
	if the factor $\frac{(b_{23}^--1)(b_{23}^-+1)(b_{23}^+-1)^2[(b_{23}^++1)u+2b_3v]u}{4b_3^2}$ is non-vanishing such that we can divide by it. Thus, a general line is sent by $g$ to a line on $\PP^3$ that passes through $p_4=[0:0:0:1]$ (when $[u:v]=[1:-1]$) and $\left[\frac{b_{23}^+-1}{b_{23}^--1} : \frac{b_{23}^++1}{b_{23}^-+1} : 1 : 0\right]$ (when $[u:v]=[1:0]$). (N.B. Again, we cannot set $u=0$.) This is a part of the del Pezzo component $D_4$ that is visible in the affine patch $U_{01,23,45}$.
	
	\item The component with $a_{24}^-=-b_{34}^+=1, a_{23}^-=b_{23}^-, (a_{23}^-)^2=-4a_2b_3+1$.
	
	If $b_3\neq0$, then we can choose the free variables to be $b_2$ and $b_3$, for which $a_2=\frac{1-(b_{23}^-)^2}{4b_3}$ and $a_3=a_2-b_{23}^-$. Then, we have:
	\begin{equation}
		\begin{bmatrix}
			u(1+a_2)+v b_2 \\
			u(1-a_2)-v b_2 \\
			u a_3+v(1+b_3) \\
			-u a_3+v(1-b_3) \\
			u(a_2-1)-v(1+b_3) \\
			-u(1+a_2)+v(b_3-1)
		\end{bmatrix}
		\xmapsto{g}
		-2v b_3 \begin{bmatrix}
			1 \\ 1 \\ 1 \\ 1
		\end{bmatrix}
		+u \begin{bmatrix}
			b_{23}^-+1 \\ b_{23}^- -1\\ -2b_3 \\ 0
		\end{bmatrix},
	\end{equation}
	if the factor $\frac{(b_{23}^--1)(b_{23}^+-1)[(b_{23}^--1)u-2b_3v]^2}{4b_3^2}$ is non-vanishing such that we can divide by it. Thus, a general line is sent by $g$ to a line on $\PP^3$ that passes through $p_5=[1:1:1:1]$ (when $[u:v]=[0:1]$) and $[b_{23}^-+1:b_{23}^--1:-2b_3:0]$ (when $[u:v]=[1:0]$). (N.B. Now there is no problem with setting $u=0$.) This is a part of the del Pezzo component $D_5$ that is visible in the affine patch $U_{01,23,45}$.

\end{enumerate}

\subsection{The affine patch $U_{02,13,45}$}
Next, consider a line on the affine patch $U_{02,13,45}$. The parameterization of a line on this patch is got by exchanging $x_1\leftrightarrow x_2$ in that of a line on $U_{01,23,45}$ to get
\begin{equation}
	U_{02,13,45}\ni x=\begin{bmatrix}
		x_0 \\ x_1 \\ x_2 \\ x_3 \\ x_4 \\ x_5
	\end{bmatrix}
	=\begin{bmatrix}
		u(1+a_2')+v b_2' \\
		u a_3'+v(1+b_3') \\
		u(1-a_2')-v b_2' \\
		-u a_3'+v(1-b_3') \\
		u a_4'+v b_4' \\
		-u(2+a_4')-v(2+b_4')
	\end{bmatrix},
\end{equation}
where, as before, $[u:v]\in\PP^1$. So, on this patch, there are the following four non-linear components:
\begin{enumerate}
	\item The component with $a_{24}'^+=b_{34}'^+=-1, a_{23}'^+=b_{23}'^+,(a_{23}'^+)^2=4a_2'b_3'+1$.
	
	If $a_2'\neq0$, then we can choose the free variables to be $a_2'$ and $a_3'$, for which $b_3'=\frac{(a_{23}'^+)^2-1}{4a_2'}$ and $b_2'=a_{23}'^+-b_3'$.
	Then, we have
	\begin{equation}
		\begin{bmatrix}
			u(1+a_2')+v b_2' \\
			u a_3'+v(1+b_3') \\
			u(1-a_2')-v b_2' \\
			-u a_3'+v(1-b_3') \\
			-u(1+a_2')-v(b_3'+1) \\
			u(a_2'-1)+v(b_3'-1)
		\end{bmatrix}
		\xmapsto{g}v\begin{bmatrix}
			1 \\ 0 \\ 0 \\ 0
		\end{bmatrix}
		+[2a_2'u+(a_{23}'^+-1)v]
		\begin{bmatrix}
			0 \\
			-\dfrac{1}{1+a_{23}'^-} \\
			\dfrac{1}{a_{23}'^+-1} \\
			-\dfrac{1}{2}
		\end{bmatrix},	
	\end{equation}
	if the factor $-\frac{(a_{23}'^--1)(a_{23}'^-+1)(a_{23}'^+-1)(a_{23}'^++1)v^2}{2a_2'^2}$ does not vanish, such that we can divide by it. Thus, $g$ sends a line to a line on $\PP^3$ passing through $p_1=[1:0:0:0]$ (when $[u:v]=[1-a_{23}'^+:2a_2']$) and $\left[0:\frac{1-a_{23}'^+}{1+a_{23}'^-}:1:\frac{1-a_{23}'^+}{2}\right]$ (when $[u:v]=[0:1]$). (N.B. We cannot set $v=0$.) This is a part of the del Pezzo component $D_1$ that is visible in the affine patch $U_{02,13,45}$.
	
	\item The component with $a_{24}'^+=-b_{34}'^-=-1, a_{23}'^-=b_{23}'^-, (a_{23}'^-)^2=-4a_2'b_3'+1$. 
	
	If $a_2'\neq0$, then we can choose the free variables to be $a_2'$ and $a_3'$, for which $b_3'=\frac{1-(a_{23}'^-)^2}{4a_2'}$ and $b_2'=a_{23}'^-+b_3'$. Then, we have:
	\begin{equation}
		\begin{bmatrix}
			u(1+a_2')+v b_2' \\
			u a_3'+v(1+b_3') \\
			u(1-a_2')-v b_2' \\
			-u a_3'+v(1-b_3') \\
			-u(1+a_2')+v(b_3'-1) \\
			u(a_2'-1)-v(b_3'+1)
		\end{bmatrix}
		\xmapsto{g}
		u
		\begin{bmatrix}
			1 \\ 1 \\ 1 \\ 1
		\end{bmatrix}
		+[2a_2'u+(1+a_{23}'^-)v]
		\begin{bmatrix}
			-\dfrac{1}{2a_2'} \\
			0 \\
			-\dfrac{1}{a_{23}'^+-1} \\
			\dfrac{a_{23}'^--1}{4a_2'}
		\end{bmatrix},
	\end{equation}
	if the factor $\frac{(a_{23}'^++1)(a_{23}'^+-1)(a_{23}'^--1)v^2}{a_2}$ does not vanish, such that we can divide by it. Thus, $g$ sends a line to a line on $\PP^3$ that passes through $p_5=[1:1:1:1]$ (when $[u:v]=[-(1+a_{23}'^-):2a_2']$) and $\left[-\frac{1}{2a_2'} : 0 : -\frac{1}{a_{23}'^+-1} : \frac{a_{23}'^--1}{4a_2'}\right] $ (when $[u:v]=[0:1]$). This is a part of the del Pezzo component $D_5$ that is visible in the affine patch $U_{02,13,45}$.
	
	\item The component with $a_{24}'^-=b_{34}'^-=1, a_{23}'^+=b_{23}'^+,(a_{23}'^+)^2=4a_2'b_3'+1$.
	
	If $a_2'\neq0$, then we can choose the free variables to be $a_2'$ and $a_3'$, for which $b_3'=\frac{(a_{23}'^+)^2-1}{4a_2'}$ and $b_2'=a_{23}'^+-b_3'$. Then, we have:
	\begin{equation}
		\begin{bmatrix}
			u(1+a_2')+vb_2' \\
			ua_3'+v(1+b_3') \\
			u(1-a_2')-vb_2' \\
			-ua_3'+v(1-b_3') \\
			u(a_2'-1)+v(b_3'-1) \\
			-u(a_2'+1)-v(b_3'+1)
		\end{bmatrix}
		\xmapsto{g}
		u
		\begin{bmatrix}
			0 \\ 1 \\ 0 \\ 0
		\end{bmatrix}
		+[2a_2'u+(1+a_{23}'^+)v]
		\begin{bmatrix}
			\dfrac{1}{a_{23}'^--1} \\
			0 \\
			\dfrac{1}{2a_2'} \\
			\dfrac{1-a_{23}'^+}{4a_2'}
		\end{bmatrix},
	\end{equation}
	if the factor $2(1-a_{23}'^-)[2a_2u+(a_{23}'^+-1)v]v$ does not vanish, such that we can divide by it. Thus, a general line is sent by $g$ to a line on $\PP^3$ that passes through ${p_2=[0:1:0:0]}$ (when $[u:v]=[a_{23}'^++1 : -2a_2']$) and $\left[\frac{a_{23}'^++1}{a_{23}'^--1} : 0 : \frac{a_{23}'^++1}{2a_2'} : \frac{(a_{23}'^-)^2-1}{4a_2'}\right]$ (when ${[u:v]=[0:1]}$). This is a part of the del Pezzo component $D_2$ that is visible in the affine patch $U_{02,13,45}$.
	
	\item The component with $a_{24}'^-=-b_{34}'^+=1,a_{23}'^-=b_{23}'^-,(a_{23}'^-)^2=-4a_2'b_3'+1$.
	
	If $a_2'\neq0$, then we can choose the free variables to be $a_2'$ and $a_3'$, for which $b_3'=\frac{1-(a_{23}'^-)^2}{4a_2'}$ and $b_2'=a_{23}'^-+b_3'$. Then, we have:
	\begin{equation}
		\begin{bmatrix}
			u(1+a_2')+v b_2' \\
			u a_3'+v(1+b_3') \\
			u(1-a_2')-v b_2' \\
			-u a_3'+v(1-b_3') \\
			u(a_2'-1)-v(b_3'+1) \\
			-u(1+a_2')+v(b_3'-1)
		\end{bmatrix}
		\xmapsto{g}
		(u+v)
		\begin{bmatrix}
			0 \\ 0 \\ 1 \\ 0
		\end{bmatrix}
		+[2a_2'u+(a_{23}'^--1)v]
		\begin{bmatrix}
			\dfrac{1}{a_{23}'^--1} \\
			\dfrac{1}{2a_2'} \\
			0 \\
			-\dfrac{(a_{23}'^+-1)}{4a_2'},
		\end{bmatrix}
	\end{equation}
	if the factor $-2(a_{23}'^--1)[2a_2'u+(1+a_{23}'^-)v]$ does not vanish, such that we can divide by it. Thus, a general line is sent by $g$ to a line on $\PP^3$ that passes through ${p_3=[0:1:0:0]}$ (when $[u:v]=[1-a_{23}'^-:2a_2']$) and $\left[1:\frac{a_{23}'^--1}{2a_2'}:0:-\frac{(a_{23}'^--1)(a_{23}'^+-1)}{4a_2'}\right]$ (when $[u:v]=[0:1]$). This is a part of the del Pezzo component $D_3$ that is visible in the affine patch $U_{02,13,45}$. 
\end{enumerate}

\bibliography{u1_squared_references}    

\end{document}